\documentclass[review]{elsarticle}

\usepackage{lineno,hyperref}
\modulolinenumbers[5]

\journal{Computer Networks}

\usepackage[nolist,withpage]{acronym}
\begin{acronym}
	\acro{SDN}{Software-Defined Networking}
	\acro{QoS}{Quality-of-Service}
	\acro{ICT}{Information and Communication Technology}
	\acro{ITT}{Inter-Transmission Time}
	\acro{HB}{Heartbeat}
	\acro{FFG}{Fast Failover Group}
	\acro{BFD}{Bidirectional Forwarding Detection}
	\acro{OF}{Open Flow}
	\acro{MPLS}{Multi Protocol Label Switching}
	\acro{SV}{Sampled Value}
	\acro{GOOSE}{Generic Object Oriented Substation Event}
	\acro{MMS}{Manufacturing Message Specification}
	\acro{SUCCESS}{Software-Defined Universal Controller for Communications in Essential Systems}
	\acro{NBI}{Northbound Interface}
	\acro{SBI}{Southbound Interface}
	\acro{JSON}{JavaScript Object Notation}
	\acro{URL}{Uniform Resource Locators}
	\acro{NC}{Network Calculus}
	\acro{IEC}{International Electrotechnical Commission}
	\acro{IED}{Intelligent Electronic Device}
	\acro{SSH}{Secure Shell}
	\acro{OVS}{Open vSwitch}
	\acro{DB}{Database}
	\acro{HTB}{Hierarchical Token Bucket}
	\acro{DFS}{Depth-First Search}
	\acro{MAS}{Multi-Agent System}
	\acro{REST}{Representational State Transfer}
	\acro{API}{Application Programming Interface}
	\acro{BSMA}{Bounded Shortest Multicast Algorithm}
	\acro{FIFO}{First In First Out}
	\acro{vSwitch}{virtual switch}
	\acroplural{vSwitch}[vSwitches]{virtual switches}
	\acro{bSwitch}{baremetal switch}
	\acroplural{bSwitch}[bSwitches]{baremetal switches}
	\acro{PFC}{Power Flow Controller}
	\acro{IP}{Internet Protocol}
	\acro{MAC}{Media Access Control}
	\acro{WAMPAC}{Wide-Area Monitoring Protection and Control}
	\acro{SCADA}{Supervisory Control and Data Acquisition}
	\acro{NIC}{Network Interface Card}
	\acro{AMR}{Automated Meter Reading}
	\acro{EV}{Electric Vehicle}
	\acro{DSM}{Demand Side Management}
	\acro{DER}{Distributed Energy Resources}
	\acro{EVSE}{Electric Vehicle Supply Equipment}
	\acro{ISO}{International Organization for Standardization}
	\acro{OCPP}{Open Charge Point Protocol}
	\acro{PLC}{Power Line Communications}
	\acro{DSL}{Digital Subscriber Line}
	\acro{TTC}{Transmission Traffic Classes}
	\acro{HVDC}{High Voltage Direct Current}
	\acro{FRR}{Fast Reroute}
	\acro{CPS}{Cyber-Physical System}
	\acro{DoS}{Denial-of-Service}
	\acro{SCD}{Substation Configuration Description}
	\acro{LLDP}{Link Layer Discovery Protocol}
	\acro{IEEE}{Institute of Electrical and Electronics Engineers}
	\acro{PES}{Power and Energy Society's}
	\acro{TLS}{Transport Layer Security}
	\acro{AI}{artificially intelligent}
	\acro{IoT}{Internet-of-Things}
	\acro{CPU}{Central Processing Unit}
\end{acronym}
\usepackage{siunitx}
\usepackage{booktabs}
\usepackage{multirow}
\usepackage{amsmath}
\usepackage{setspace}
\usepackage[ruled,linesnumbered]{algorithm2e}
\usepackage{array}
\usepackage[pstex,natural,dvipsnames]{xcolor}
\usepackage{color}
\usepackage{colortbl}
\usepackage{graphicx}
\usepackage{subfig}
\begin{document}

\begin{frontmatter}

\title{Enabling Hard Service Guarantees in\\Software-Defined Smart Grid Infrastructures}
\tnotetext[mytitlenote]{\copyright 2018. This manuscript version is made available under the CC-BY-NC-ND 4.0 license http://creativecommons.org/licenses/by-nc-nd/4.0/.
	The formal version of this publication is available via \href{https://doi.org/10.1016/j.comnet.2018.10.008}{10.1016/j.comnet.2018.10.008}.
}

\author{Nils Dorsch\corref{mycorrespondingauthor}}
\cortext[mycorrespondingauthor]{Corresponding author}
\ead{nils.dorsch@tu-dortmund.de}
\author{Fabian Kurtz}
\author{Christian Wietfeld}
\address{TU Dortmund University\\Otto-Hahn-Str. 6, 44227 Dortmund, Germany}

%
%

\begin{abstract}
\ac{ICT} infrastructures play a key role in the evolution from traditional power systems to Smart Grids.
Increasingly fluctuating power flows, sparked by the transition towards sustainable energy generation, become a major issue for power grid stability.
To deal with this challenge, future Smart Grids require precise monitoring and control, which in turn demand for reliable, real-time capable and cost-efficient communications.
For this purpose, we propose applying \ac{SDN} to handle the manifold requirements of Smart Grid communications.
To achieve reliability, our approach encompasses fast recovery after failures in the communication network and dynamic service-aware network (re-)configuration.
\ac{NC} logic is embedded into our \ac{SDN} controller for meeting latency requirements imposed by the standard \acs{IEC}~61850 of the \ac{IEC}.
Thus, routing provides delay-optimal paths under consideration of existing cross traffic.
Also, continuous latency bound compliance is ensured by combining \ac{NC} delay supervision with means of flexible reconfiguration.
For evaluation we consider the well-known Nordic 32 test system, on which we map a corresponding communication network in both experiment and emulation.
The described functionalities are validated, employing realistic \ac{IEC}~61850 transmissions and distributed control traffic. 
Our results show that hard service guarantees can be ensured with the help of the proposed \ac{SDN} solution.
On this basis, we derive extremely time critical services, which must not be subjected to flexible reconfiguration.
\end{abstract}


\end{frontmatter}


\section{Introduction}
\label{sec:intro}
%
%
%
%

\acresetall

Future power systems are faced with severe challenges, caused by the transition from conventional to distributed, renewable generation \cite{smartgrid_survey}. 
To fully exploit the advantages and mitigate the drawbacks of fluctuating power generation from these energy resources, concepts such as \ac{DSM} and controllable loads/storages, e.g. scheduling \ac{EV} charging, need to be applied.
At the same time, the energy system has to deal with further volatile power transmissions, caused by increasing energy trade due to the liberalization of energy markets.
Resulting from these challenges, precise monitoring and control of the system are indispensable for maintaining grid stability and avoiding cascading outages.
Subsequently, appropriate \ac{ICT} infrastructures are required to ensure reliable, timely transfer of measurement data and control commands, in particular on transmission grid level \cite{smartgrid_comm_survey,NAFI201623}. 
Quantitative requirements are given in the \ac{IEC}'s standard \ac{IEC}~61850, which is set to become the prevailing normative for power grid communications.
It defines intervals as low as \SI{250}{\mu s} and maximum allowed latencies of \SI{5}{ms} for measurement data transmission and protection tripping respectively \cite{IEC61850:TC57}.
Meanwhile, distribution grid communications deal with numerous protocols and a variety of different access technologies \cite{albano_lastmile}.
Overall, an increasing number of \ac{IED}, each with distinct service requirements, will be connected to wide area communication networks.

To cope with these specific demands of Smart Grid communications, we propose a comprehensive framework, building on the concept of \ac{SDN}.
In this way, we are able to provide hard service guarantees with traffic flow granularity.
\ac{SDN} constitutes a promising new take on networking, offering flexible, dynamic configuration of communication infrastructures \cite{ofwhitepaper}. 
Following the paradigm of separating data and control planes, \ac{SDN} establishes a programmable controller platform.
It enables managing traffic flows, profiting from a global network view.
There exist various mechanisms for enhancing particular aspects of communications' \ac{QoS}.
Yet, they typically suffer from vendor specific peculiarities, poor integration and overly complicated configuration \cite{sydney2013simulative}.
In contrast, our approach is able to address the multitude of diverging requirements, while allowing for straightforward extension and configuration.
In particular, this concept provides means for fast failure recovery, dynamic prioritization and queue configuration under the overall paradigm of application- and \ac{QoS}-awareness.
\ac{NC} algorithms \cite{le_boudec_network_2004} are incorporated into our \ac{SDN} controller to predict and monitor end-to-end delays of traffic flows analytically.
Hence, violations of delay bounds can be identified in time to activate counter-measures, ensuring continuous fulfillment of hard real-time guarantees. 

\begin{figure}[t]
	\centering
	\includegraphics[width=0.9\columnwidth]{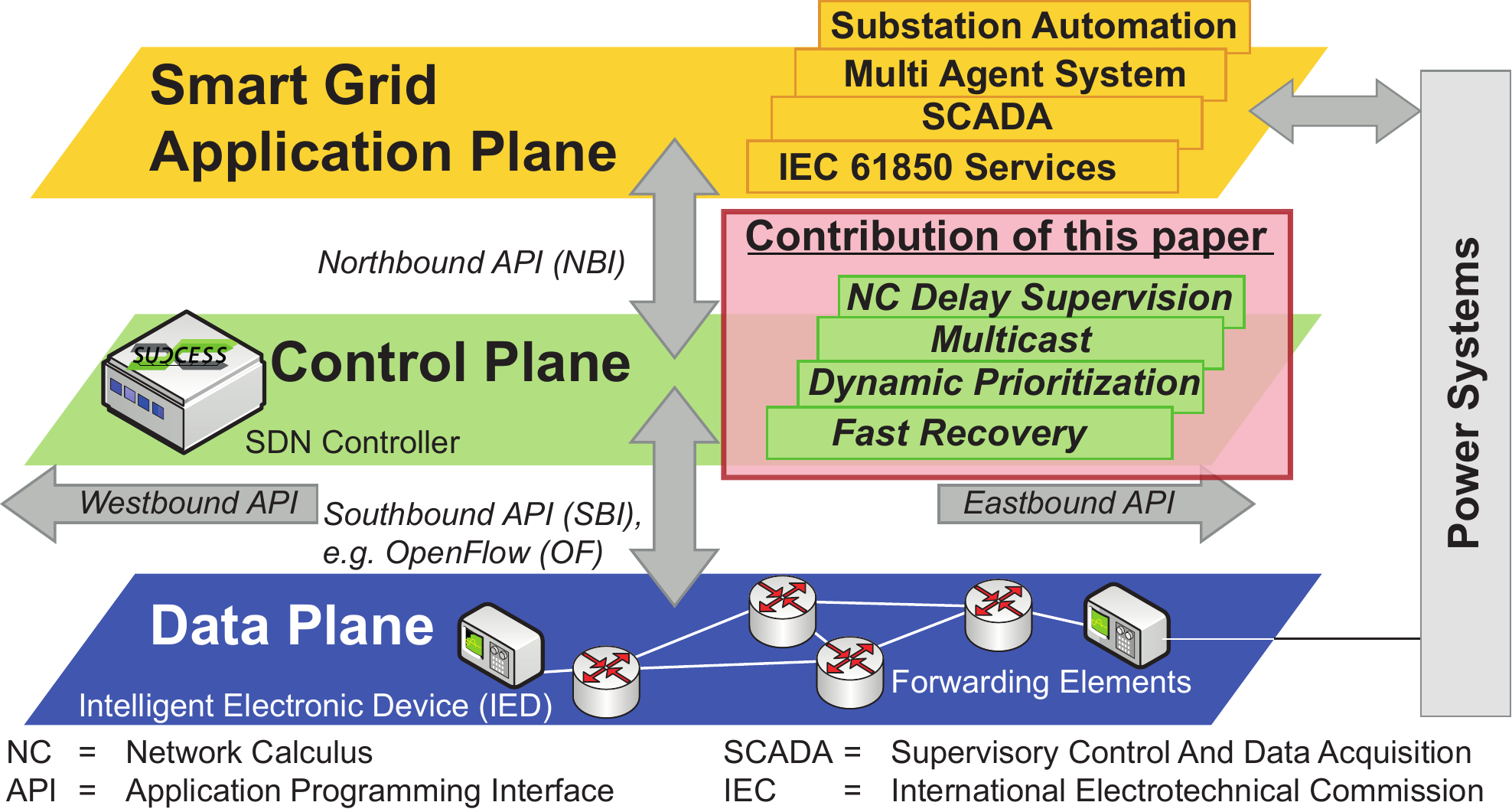}
	\caption{Solution approaches addressed in this paper, mapped on the Software-Defined Networking for Smart Grids concept, introduced in \cite{SDN4SG}}
	\label{fig:sdn_arch} 
\end{figure}

The main contributions of this paper are the following:
\begin{itemize}
	\item \acl{SDN} enabled service-centric network configuration and adaption for Smart Grids, providing hard service guarantees
	\item the integration of \ac{NC} into \ac{SDN}-driven network control for delay supervision and routing to ensure real-time capable communications at all times
\end{itemize}
Figure \ref{fig:sdn_arch} provides an overview of our concept for \ac{SDN}-enabled Smart Grid communications, highlighting interactions between \ac{ICT} and power system applications.
We evaluate our concepts, considering \ac{IEC}~61850 communications as well as a \ac{MAS} for distributed control on a fiber-based communication infrastructure for the Nordic 32 test system \cite{nordic32}.
Both empirical measurements and emulations of the whole infrastructure are utilized.
In addition, the proposed concepts may be adapted to other mission critical systems such as transportation or rescue services.

This work has been carried out as part of larger scale research efforts, i.e. DFG research unit 1511 and the Franco-German project BERCOM.
Subsequently, Smart Grid requirements were synchronized and solution approaches discussed with power system experts and utilities such as EDF.

The remainder of this work is structured as follows:
Section~\ref{sec:stateart} provides an overview of the state-of-the-art, detailing the requirements of Smart Grid communications and introducing the main principles of \ac{SDN} and \ac{NC}.
The section is completed by an overview of related work.
Next, we describe our solution approach based on the \ac{SDN} controller framework (Section~\ref{sec:approach}).
In Sections~\ref{sec:scenario} and \ref{sec:testbed} a description of the Smart Grid scenario and an overview of the developed testbed set-up are provided.
Afterwards, empirical, emulation and analytical evaluation results are presented in Section~\ref{sec:results}.
Finally, the paper concludes with a summary and an outlook on future work (Section~\ref{sec:conclusion}).


\section{State-of-the-Art on Smart Grid Communications, Software-Defined Networking and Associated Performance Evaluation}
\label{sec:stateart}
This section reviews Smart Grid communication requirements and reflects on the state-of-the-art of \acf{SDN} and \acl{NC}, for enabling respectively verifying hard service guarantee compliance.
Afterwards, results of related work are described and compared to this article.

\subsection{Smart Grid Communication Use Cases}
\label{sec:usecases}
Smart Grid communication requirements can be roughly divided into distribution and transmission grid use cases, as detailed below.
While these power system levels exhibit widely diverging demands, \ac{SDN} offers an integrated approach for associated communications.

\subsubsection{Managing the Distribution Power Grid}
Communication-dependent applications in the distribution power grid comprise \ac{AMR}, \ac{DSM}, monitoring and control of \ac{DER}, as well as coordination of \ac{EV} charging.
\ac{AMR} is considered a fundamental function of smart distribution grids, providing measurement data as the basis for more advanced applications, such as novelty detection power meters \cite{nodepm}. 
For this concept machine learning is deployed on distributed energy measurement data to optimize the energy consumption times of end users.
Also, anomalies can be detected, revealing energy consumption that deviates from common patterns (e.g. non-technical losses).
This concept can be further enhanced by integrating an intelligent decision-making system for reducing energy consumption on basis of temporal correlations \cite{Filho2018ResiDITA}.
High precision decision-making is achieved with the help of artificial neural networks. 
Such approaches mark the transition to \ac{AI} energy systems, focused on energy efficiency, providing an evolution of \ac{DSM}.

Design and operation of \ac{ICT} infrastructures for the distribution power grid are driven by large numbers of devices, heterogeneity of protocols and technologies \cite{KUZLU201474}.
While \ac{IEC}~61850 becomes increasingly important for \ac{DER} control, dedicated sets of protocols are applied for \ac{AMR} (e.g. IEC~62056, DLMS/COSEM) and \ac{EV} charging (e.g. ISO~15118 and OCCP).
For physical transmission, various wired (\ac{PLC}, broadband cable) and wireless access technologies (WiFi, cellular) are considered.
Moreover, driven by business-to-consumer use cases, aspects like role management, authentication and billing play an important role.

\subsubsection{Controlling the Transmission Power Grid}
In contrast to distribution systems, communications on the transmission grid level focus on requirements such as reliability, real-time capability and security.
Use cases involve substation automation including extremely time critical protection functions, \ac{WAMPAC} and \ac{SCADA}.
Fiber-optic infrastructures are regarded as main transmission medium, whereas cellular networks are considered as alternative or back-up solution for the network access domain.

\begin{table}[h]
	\centering
	\caption{Smart Grid timing requirements, specified in \ac{IEC}~61850-5 \cite{IEC61850:TC57}}
	\begin{tabular}{c c c}
		\toprule
		\parbox[t]{1.3cm}{\centering \textbf{Transfer\\Time Class}} & \parbox[t]{2.3cm}{\centering \textbf{Maximum}\\\textbf{Transfer Time [ms]}}& \textbf{Type of Transfer}\\
		\midrule
		0 & $>$ 1000 & files, events, logs \\
		\midrule
		1 & 1000 & events, alarms \\
		\midrule
		2 & 500 & operator commands \\
		\midrule
		3 & 100 & slow automatic interactions \\
		\midrule
		4 & 20 & fast automatic interactions  \\
		\midrule
		5 & 10 & releases, status changes \\
		\midrule
		6 & 3 & trips, blockings \\	
		\bottomrule
	\end{tabular}
	\label{tab:timereq}
\end{table}

\paragraph{Centralized Power System Control}
\ac{SCADA} provides the basis for centralized grid control functionalities.
Protocol-wise \ac{IEC}~60870 is currently still widely applied for this purpose.
However, \ac{IEC}~61850, originating from substation automation, is about to become the dominating protocol throughout transmission system communications (as well as for some distribution grid applications).
It employs a holistic approach, covering detailed data models for devices and functions, abstract communication service descriptions as well as actual protocols.
Measurement values are transmitted in fixed intervals of \SI{250}{\mu s}, using \ac{SV} messaging.
The \ac{GOOSE} service is applied for exchanging statuses and issuing switching commands. 
Both message types are encapsulated into Ethernet packets directly.
\ac{GOOSE} operates semi-regularly with periodic status messages in intervals of e.g. \SI{1}{s}, whereas commands are issued in response to events and are repeated in increasing intervals starting at \SI{1}{ms}.
Meanwhile, \ac{MMS} utilizes client-server-based TCP/IP communication for tasks like software updates, configuration and measurement reports.  
Table \ref{tab:timereq} provides an overview of end-to-end timing demands for different applications in \ac{IEC}~61850, regardless of communication failures.
The requirements are divided into corresponding \ac{TTC}, defining maximum transfer times \cite{IEC61850:TC57}.

\paragraph{Distributed Power System Control}
Differing from the common \ac{SCADA} approach, power systems may also be controlled in a distributed manner, utilizing for example a \acf{MAS}.
Such an \ac{MAS} is introduced in \cite{muller2014}, placing agents at substations of the power grid.
These agents utilize local information along with data from adjacent substations, received via inter-agent communication, to gain an estimate of the surrounding power grid's state.
In case emergency conditions are detected, the agents coordinate counter-measures and apply local assets to stabilize voltage and prevent black-outs.
For example, set points of \ac{HVDC}-converters and power flow controllers can be changed.
Also, re-dispatch of flexible generation and load may be initiated. 
A first integration between a JAVA-based implementation of this distributed grid control and our \ac{SDN} controller framework was achieved in \cite{dorsch_intertwined}.

\subsection{Software-Defined Networking Enabled Communication Systems}
\label{sec:sdnconcept}
Software-Defined Networking is a novel approach towards networking, based on the idea of separating control and data plane \cite{ofwhitepaper}.
Therefore, control functionalities are abstracted from networking nodes and consolidated at a dedicated instance, known as the \ac{SDN} controller.
Hence, data plane devices become \ac{SDN} switches, handling physical transmission of packets only.
Unknown traffic flows are forwarded to the \ac{SDN} controller for classification.
This central component handles routing and installs corresponding forwarding rules at all relevant devices throughout the network.
Subsequent packets of the same traffic flow are handled by the data plane components on basis of the rules established previously.
Communication between the \ac{SDN} controller and the forwarding elements is handled via the so-called \ac{SBI} with \ac{OF} \cite{ofspec} being the most prominent -- de-facto standard -- protocol for this purpose \cite{FARHADY201579}.

One major benefit of \ac{SDN} is the controller's programmability, which -- in conjunction with its global network view -- can be used to adapt dynamically to changes in the communication network.
Moreover, it allows for straightforward integration of a variety of different approaches and algorithms, like for example traffic engineering capabilities of \ac{MPLS}.
While integrating such functionalities, \ac{SDN} obviates overly complex configuration, usually associated with such approaches \cite{sydney2013simulative}. 
Thus, network management and control are simplified significantly. 
Through its \ac{NBI} the \ac{SDN} controller discloses means of conveying communication requirements and influencing network behavior to external applications.
Contrary to the \ac{SBI}, there is no common protocol for the \ac{NBI}, though the \ac{REST} \ac{API} is in widespread use \cite{sdnsurvey}.
To achieve scalability of the \ac{SDN} approach, i.e. for controlling large infrastructures, interaction with other controllers and legacy networks is enabled via the westbound and eastbound interface respectively.

Today, \ac{SDN} is already widely deployed in data centers of companies such as Alphabet/Google \cite{google_sdn} and is considered as the foundation for communications in the core of 5G mobile communication networks \cite{sdn_5g}.

\subsection{Network Calculus for the Performance Evaluation of\\Communication Infrastructures}
\label{sec:ncprincples}
To obtain a precise, real-time view on the delay of Smart Grid communications, \ac{NC} is integrated into the controller framework as an analytical modeling approach for delay computation.
\ac{NC}, originating from the initial works of Cruz~\cite{cruz_calculus_1991} in the early 1990s, is a well-established method for the worst-case analysis of communication networks.
It is suited for arbitrary types of traffic as the approach is agnostic to statistical distribution functions, providing performance bounds only.
Current advancements of \ac{NC} favor the use of tighter, stochastic bounds, which come at the price of small violation probabilities \cite{fidler_nc}.
In this work, however, the original, deterministic \ac{NC} is applied, as timing requirements of communications in transmission power grids are extremely strict and violations may result in a fatal collapse of the system.
Hence, thorough, deterministic delay bounds, excluding any violations, are considered most suitable.

Originating from \ac{NC} terminology, we introduce \textit{flow-of-interest} and \textit{cross traffic flows} as major terms for describing network behavior in this article.
\begin{itemize}
	\item \textbf{Flow-of-interest} refers to the packet transfer, which is in the current focus of analysis.
	\item \textbf{Cross traffic flows} are other transmissions that are concurrently active on the same network and may interfere with the flow-of-interest.
\end{itemize}
To model traffic, arriving to the communication system, we employ the frequently used, leaky (token) bucket arrival curve in Equation \ref{eq:nc_arrcurv}.

\begin{equation}
\label{eq:nc_arrcurv}
\alpha(t)=\sigma+\rho\cdot t,
\end{equation}

where $\sigma$ is the maximum packet size and $\rho$ the sustained date rate requirement of the traffic flow.
These parameters follow pre-defined values per assigned traffic/priority class.
To map the service, which is offered to the traffic flow by network elements such as links or switches, the concept of service curves is adopted.
Here, we use rate latency curves per outgoing switch port, considering data rate $R$ and propagation delay $T_{pr}$ of the link as well as transmission ($T_{tr}$) and switching delay ($T_{sw}$):

\begin{equation}
\beta(t)=R\cdot[t-T]^+,
\end{equation}

with $T=T_{pr}+T_{tr}+T_{sw}$. 
By linking arrival and service curves, the delay and backlog, that is experienced by the flow-of-interest at the respective network element, can be determined.
To obtain the traffic flow's overall network delay bound directly, \ac{NC} utilizes the concept of the end-to-end service curve.
It is calculated as the convolution of all service curves on the flow's path, as given by Equation \ref{eq:nc_etesc}.

\begin{equation}
\label{eq:nc_etesc}
\beta_{end-to-end,i}(t)=\beta_{1,i}(t)\otimes...\otimes\beta_{n,i}(t),
\end{equation}

with $1...n$ being the index of the switches on the path between source and destination.
The interference of other transmissions, \textit{cross-traffic flows}, is captured by the left-over-service curve $\beta_{k,i}(t)$ with $i$ being the index of the flow-of-interest and $k$ identifying the respective switch.
It is defined by Equation \ref{eq:nc_losc} and describes the service, which can still be provided to the flow-of-interest after taking into account interfering traffic.

\begin{equation}
\label{eq:nc_losc}
\beta_{k,i}(t)=\beta_{k,base_{i}}(t)-\sum^{m}_{j=i}\left(\alpha_{k,j}(t-\Theta)\right),
\end{equation}

where cross traffic flows of same or higher priority ($j=i...m$) reduce the service available to flow $i$.
Subsequently, the cross traffic arrival curves $\alpha_{k,j}$ of flow $j$ at node $k$ are subtracted from the specific base service curve of flow $i$.
For flows of higher priority ($j>i$) strict prioritization is assumed, resulting in $\Theta=0$, whereas for flows of the same priority \ac{FIFO} scheduling applies, introducing $\Theta$ as additional level of flexibility.

\subsection{Related Work}
\label{sec:related}
In recent years, \ac{SDN} has been a major topic of research with numerous related publications.
Hence, our review focuses on a subset of these works, i.e. papers which apply \ac{SDN} in the context of Smart Grids or aim at integrating \ac{SDN} with \ac{NC}.

Starting with the latter, Guck \textit{et al.} split online routing and \ac{NC}-based resource allocation, achieving average link utilization close to the results of mixed-integer programming in software-defined industrial \ac{ICT} infrastructures \cite{guck_sdnnc}.
In contrast to our approach, performance is assessed individually for each node, instead of applying end-to-end bounds, which are known to be tighter \cite{fidler_nc}.
\ac{NC} is applied in \cite{duan_sdnnc} to create a high-level abstraction model of network service capabilities, guaranteeing inter-domain end-to-end \ac{QoS}.
Thus, the authors derive the required bandwidth of services, whereas this work focuses on end-to-end latency guarantees. 
In \cite{qin_sdnnc} a variation of \ac{NC} serves as basis for a multi-constraint flow scheduling algorithm in \ac{SDN}-enabled \ac{IoT} infrastructures.
The performance of \ac{SDN} deployments is evaluated, modeling \ac{SDN}~controller-switch interactions with \ac{NC} in \cite{sdndeploy_nc}.
Yet, computations are performed offline as the approach is not coupled with an actual \ac{SDN} set-up.
Similarly, Huang \textit{et al.} validate their proposed hybrid scheduling approach for \ac{SDN} switches by applying offline \ac{NC} analysis \cite{HUANG201789}.
In \cite{KHORSANDIKOOHANESTANI201731} \ac{NC} is employed for the analysis of \ac{SDN} scalability.
Therefore, the authors determine worst case delay bounds on the interaction between network nodes and \ac{SDN} controller.
The approach considers switch internals and utilizes similarities between flow tables and caches.
Evaluations indicate sensitivity to parameters such as network and flow table size, traffic characteristics and delay, allowing to deduce recommendations for distributed controller concepts.
Just as the previous two articles, publication \cite{KHORSANDIKOOHANESTANI201731} analyzes \ac{SDN}-enabled infrastructures with the help of \ac{NC}, but does not integrate it with the system.

In previous studies we modeled a traditional wide-area communication network for transmission systems on basis of \ac{IEC}~61850 and evaluated its real-time capability using \ac{NC} \cite{nc_infocom}.
The developed framework serves as a starting point for combining \ac{NC} and \ac{SDN} within this article.

A general overview of possible applications of \ac{SDN} in \acp{CPS} is given in \cite{MOLINA2018407}.
With regard to Smart Grid communications, Cahn \textit{et al.} proposed \ac{SDN}-based configuration of a complete \ac{IEC} 61850 substation environment \cite{cahn2013software}.
Molina \textit{et al.} propose an \ac{OF}-enabled substation infrastructure, integrating \ac{IEC}~61850 configuration into the Floodlight controller by reading \ac{SCD} files \cite{MOLINA2015142}.
In this way, the approach is very similar to the concepts presented in \cite{cahn2013software}.
Based on the configuration file, static traffic flows with different priorities are established.
Mininet is employed to test functionalities such as traffic prioritization, detection of \ac{DoS} attacks and load balancing.
However, these use cases show only minor advancements compared to standard Floodlight, whereas the main contribution is automatic substation network configuration.
In \cite{dasilva_sdnscada} \ac{SDN} is utilized to design a network intrusion detection system for \ac{SCADA} communications.
To facilitate the communication between smart meters and the control centers, aggregation points are introduced to the \ac{SDN} data plane in \cite{wang_applan}.
Planning of these is optimized with respect to minimal costs applying a mathematical model.
In \cite{REN20181251} \ac{SDN} is used for establishing networked microgrids, enabling event-triggered communication.
According to the authors, in this way costs are reduced, while system resilience is enhanced.
The above publications illustrate specific use cases of \ac{SDN} in Smart Grids and are included in this literature review mainly to illustrate the broad scope of possible applications.

Sydney \textit{et al.} compare \ac{MPLS}- and \ac{OF}-based network control for power system communications, demonstrating that \ac{SDN} achieves similar performance, while simplifying configuration \cite{sydney2013simulative}.
The authors expanded their work by experiments on the GENI testbed \cite{SYDNEY20145}.
Evaluations are performed using the example of demand response, where load shedding is triggered to maintain frequency stability.
In this context, three functionalities are tested: fast failover, load balancing and \ac{QoS} provisioning.
Thus, the paper addresses topics quite similar to this article.
However, no standard Smart Grid communication protocol is applied.
Also, the publication is rather focused on the electrical side, whereas some communication aspects are not studied in full detail. 
For example, the presented recovery process is comparably slow with delays of up to \SI{2}{s} and would require further optimization.
In addition, our investigation considers further functionalities such as dynamic network reconfiguration and delay supervision.
Mininet emulation, integrated with ns-3 simulation, is used in \cite{aydeger_sdnrecov} to evaluate \ac{SDN}-based failure recovery to wireless back-up links in a Smart Grid scenario.
\ac{OF} \ac{FFG} are used in \cite{pfeiffenberger_sdnreliability} to enable fault-tolerant multicast in Smart Grid \ac{ICT} infrastructures.
Both of the above papers tackle specific aspects of reliability in terms of fault-tolerance, which are not addressed in this work (utilization of wireless back-up paths and multicast recovery). 
Although, the discussed papers are limited to particular realizations of fault-tolerance concepts, they could provide valuable extensions of this work. 
In contrast, this work considers reliability in a broader sense, considering the fulfillment/enforcement of data rate and latency guarantees.

In previous work we proposed an \ac{SDN} controller framework, which provides fault tolerance and dynamically adaptable service guarantees for Smart Grid communications \cite{SDN4SG,dorsch_intertwined,GC2016}.
Compared to these publications and other related work discussed above, we achieve the following improvements and contributions in this paper:
\begin{itemize}
	\item comprehensive comparison of different fast recovery approaches, quantifying path optimality and detection overhead in addition to recovery delays
	\item delay impact of dynamic network reconfiguration in response to Smart Grid service requirements and network conditions, illustrated on a five step sequence of events
	\item delay-aware routing using \ac{NC}
	\item compliance to hard service guarantees on basis of \ac{NC} delay supervision
\end{itemize}

\section{Proposed Solution Approach for Smart Grid Communications on Basis of Software-Defined Networking}
\label{sec:approach}
To address the challenges of communications in critical infrastructures such as the Smart Grid, we propose the \ac{SUCCESS}\footnote{The source code of \ac{SUCCESS} is publicly available via https://gitlab.kn.e-technik.tu-dortmund.de/cni-public/success}.
It is a Java-based framework, designed to meet hard service requirements of mission critical infrastructures.
The framework was forked from the open-source Floodlight controller \cite{floodlight} and utilizes OpenFlow v1.3 \cite{ofspec}.

\begin{figure}[t]
	\centering
	\includegraphics[width=\columnwidth]{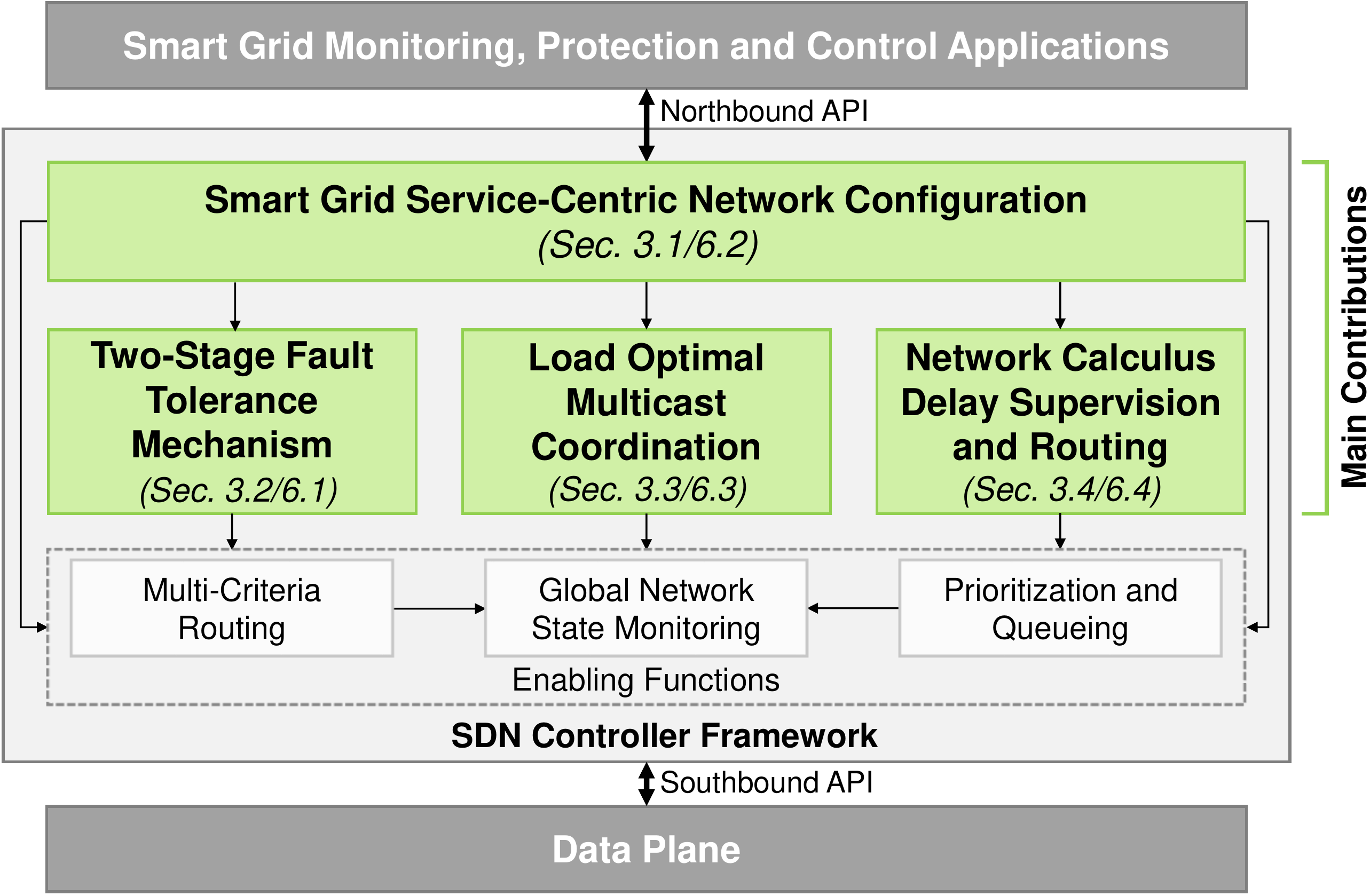}
	\caption{Elements of the \acl{SUCCESS}, their interdependencies and classification within the \ac{SDN} concept (including reference to corresponding discussions)}
	\label{fig:success} 
\end{figure}

Figure \ref{fig:success} illustrates the different components of our controller, including their interdependencies as well as the connection to Smart Grid applications via the \acf{NBI}.
As a basis for the main contributions of this work, we devise the following functions:
\begin{itemize}
	\item \textit{Global Network State Monitoring}: Active traffic flows as well as link states are tracked to obtain a real-time view of the current network load.
	\item \textit{Multi-Criteria Routing}: In contrast to standard optimal path routing, we employ \ac{DFS} to determine multiple feasible routes, which can be applied as alternatives for fast failure recovery and hard service guarantee provisioning.
	\item \textit{Prioritization and Queuing}: For prioritization we apply a large range of priority levels, which are mapped to corresponding queues, which encompass minimum and maximum data rate guarantees on basis of Linux \ac{HTB} \cite{htb}. 
	
	We enable controller-driven, flexible queue configuration by modifying \ac{OVS} \ac{DB} entries with the help of \ac{OVS} commands via \ac{SSH}.
	Our \ac{SDN} controller includes a dedicated module for establishing and handling \ac{SSH} sessions.
	To avoid the overhead of repeated handshake processes, sessions are maintained and provided for reuse. 
	According to our measurements the configuration of new queues incurs a mean delay of \SI{273}{ms} (\SI{601}{ms} if the \ac{SSH} session needs to be established).
	For the dynamic adaption of Smart Grid service requirements (c.f. Section \ref{sec:approach:nbi}), switching between existing queues is utilized.
	Hence, queue re-configuration is not considered time-critical.
\end{itemize}

\subsubsection*{Control Plane Considerations}
In the following, we refer to the control plane as a single instance.
However, we acknowledge the need for deploying distributed or hierarchical systems of multiple controllers for large-scale real world scenarios.
To achieve real-time reconfiguration of communication networks in such scenarios, utilizing multiple controllers to manage defined network partitions is inevitable \cite{controlplane}. 
Vice versa, in real-world scenarios, relying on a single controller induces the following issues:
First, extending the network size would result in increasing numbers of flows to be handled by the controller.
This could lead to increased calculation times and, in case of long transmission distances, to higher delays in the distribution of \ac{SDN} controller commands.
In the worst case, the controller might be overloaded completely.
With regard to the proposed \ac{NC} routing and supervision, high numbers of flows might also compromise the feasibility of the whole approach, if computing times exceed Smart Grid delay requirements.
To this end, the scalability analyses in Section \ref{sec:nc:eval:opt} may indicate network partition sizes suitable for our approach.
Yet, it will need to be assessed how traffic flows, traversing the domains of multiple controllers can be handled by NC routing and delay supervision.
Possible approaches include exchanging intermediate calculation results or the summation of delay bounds, both building upon inter-controller communication.
Also, measurement values may be integrated for this purpose.

Second, architectures with only one controller would generate a single-point-of-failure with regard to reliability and security.
If the controller or the route to it fails or is compromised by an attacker, switches can fall back to a simple layer-2 operation mode \cite{ofspec}.
However, all desired features such as hard service guarantees or the routing of new flows would be suspended.
Nevertheless, as inter-controller coordination represents an entire research area of its own, we consider it out-of-scope for this work.
Though, in another publication we discuss this topic with respect to control plane reliability \cite{kurtz_controlplane_reliable}.

Control plane networks are classified as either in- or out-of-band control.
For our experiments, we apply out-of-band control, utilizing dedicated network links to each switch.
Yet, for real-world deployments, in-band control may be better suited, as no second, parallel communication infrastructure needs to be established.
In-band control may for example be realized as internal flows of higher priority \cite{ovs}.
Despite the fact that the peculiarities of in-band control are not evaluated in this work, we would like to stress some important preconditions:
\begin{itemize}
	\item To ensure reliable transmission of control traffic, the controller must be connected to the data network via multiple links, protected by fast failover mechanisms.
	\item Control traffic needs to be estimated beforehand and kept to a minimum.
	 Thereby, the network's capacity can mostly be allocated to actual data traffic.
	\item It has to be ensured that data and control traffic do not interfere with each other, for example by using dedicated queues with appropriate priorities.
\end{itemize}

\subsection{Smart Grid Service-Centric Network Configuration}
\label{sec:approach:nbi}
For adapting communication network configurations to Smart Grid specific requirements, we enable power system applications to convey their demands to the controller.
Therefore, we implement the \ac{SDN} \ac{NBI}, using the \ac{REST} \ac{API}.
While the controller is set up as the \ac{REST} server, applications act as clients, sending requests to the controller.
Interaction via the \ac{NBI} is demonstrated employing the \ac{MAS} as client application.
Four different services -- Rule Creation, Route Reservation, Flow Modification, Multicast Group Creation and their respective revocations -- are provided by the controller.
Details on these \ac{NBI} services are provided below.

\subsubsection{Rule Creation} 
Rule Creation serves to register traffic flows at the controller, disclosing their specific demands regarding minimum data rate, maximum latency and packet loss as well as priority.
This information is stored at the controlled as combined flow requirements.
Thus, incoming traffic can be routed and directed to an adequate priority queue, fulfilling its requirements.
Hence, this functionality relies heavily on the routing, prioritization and queuing mechanisms, described previously.
Applying the DELETE command in conjunction with Rule Creation removes the respective traffic rule.

\subsubsection{Route Reservation}
Typically, in \ac{SDN}-enabled infrastructures network devices contact their associated controller to request routes for newly arriving packet streams.
This incurs additional delay for the first packets of a transmission.
Route Reservation, however, is applied to route traffic flows and configure flow table entries in advance, avoiding this initial delay.
However, such static flow table entries need to be removed explicitly, since idle time-outs are precluded.

\subsubsection{Flow Modification}
Existing flow requirements, involving priority and queue assignments, may be altered using this request.
Hence, it becomes possible to raise or reduce flow priorities temporarily, e.g. in response to emergency situations.
In particular, this request may be performed in case of simultaneous overloads of power and communication system.
Thus, successful transmission of critical commands for relieving the power grid crisis can be ensured.
Temporary changes to the flow requirements can be revoked with the help of the corresponding DELETE command. 

\subsubsection{Multicast Group Creation} 
We provide dedicated \ac{NBI} requests, enabling Smart Grid applications to trigger generation, modification and deletion of multicast groups.
To create a new multicast group, the controller is supplied with a list of \ac{MAC} or \ac{IP} addresses, representing member devices.
In addition, a set of header fields defines the messages, applicable for multicast transmission.
Hence, the controller is able to identify multicast packets and determine appropriate routes to all destinations.
The use of specific multicast addresses is not required.

\subsubsection{Security Considerations}
Though not within the scope of this work, we acknowledge the fact that securing interactions between controller, switches and applications is of critical importance. 
For mutual authentication on the switch-controller interface, \ac{OF} provides \ac{TLS} \cite{ofspec}. 
Similarly, for real-world application of our proposed northbound interface implementation, \ac{TLS}-protected communication is required. 
Additionally, our concept accounts for future security enhancements such as authentication and permission systems to ensure legitimate access \cite{hayward_security}.
Otherwise, attackers could inflict damage by requesting:
\begin{itemize}
	\item Unsuitable traffic flow configurations.
	For example, the priority and traffic demands of a single flow could be increased to a level that suppresses other data streams.
	Vice versa, flow parameters of critical Smart Grid transmissions could be manipulated to destabilize the power system.
	\item Fake multicast groups could be established to forward traffic to unauthorized parties. 
\end{itemize}

\subsubsection{Further Aspects of Smart Grid Adaptation}
Besides the aforementioned means of direct participation, \ac{SDN} provides further benefits, facilitating Smart Grid communications.
As \ac{IEC}~61850 is becoming a comprehensive standard for power systems, its application for wide area communications is discussed.
Technical reports propose the transmission of Ethernet-based \ac{SV} and \ac{GOOSE} messages over \ac{IP} systems, necessitating tunneling or conversion of packets to routableGOOSE/routableSV \cite{61850-90-5:TC57:IEC,61850-90-12:TC57:IEC,61850-90-1:TC57:IEC,61850-90-2:TC57:IEC}.
In contrast, packet routing and forwarding in \ac{OF}-enabled infrastructures builds on matches -- sets of arbitrary header fields -- and thus is protocol-agnostic.
This allows for direct transmission of \ac{IEC}~61850 \ac{SV} and \ac{GOOSE} messages on wide area networks.

\subsection{Two-Stage Fault Tolerance Mechanism}
\label{sec:approach:failover}
Guaranteeing reliable, virtually uninterrupted, transmission is a major requirement for mission-critical communications.
Therefore, mechanisms enabling fast recovery after link failures are integrated into the controller.
Failover can be split into two steps: failure detection and traffic restoration.
Both functions can be realized either locally at the switches or centrally, triggered by the \ac{SDN} controller.
To leverage the advantages of central and local algorithms at the same time, we unify both approaches to obtain a straightforward two-stage hybrid solution.

Besides complete link failures, networks may experience partial/intermittent link disruptions or high packet loss as results of malfunctioning hardware.
Depending on the selected sensitivity (i.e. detect multiplier, time-out interval and \ac{ITT}), link failure detection may discover recurring link disturbances as well.
Nonetheless, such configurations may lead to false positives.
For identifying packet loss, on the other hand, we apply \ac{OF} statistics collections.
In case the number of packets lost within the collection period exceeds a predefined threshold, traffic is redirected to alternative paths, similar to link failure recovery.
However, due to the associated higher traffic load, such statistics collections are typically performed in intervals of several seconds.
Hence, detecting packet loss is considerably slower than link failure detection.
Overall, if faulty hardware is identified, traffic may be switched to alternative paths, avoiding the affected equipment.
Yet, as described above, fast detection of phenomena such as high packet loss or intermittent link behavior is more challenging compared to the complete failure of entire links.
Eventually, such incidences may endanger latency guarantees.

\paragraph{BFD-based Local Recovery}
\ac{BFD} \cite{katz_bfd} is deployed to reduce failure detection times locally at the switches.
It is integrated into \ac{OVS} since version 2.3.0 \cite{ovs} and is also applied in combination with \ac{MPLS} \ac{FRR} to achieve fast recovery in \ac{MPLS}-based infrastructures \cite{mpls_bfd}.
For monitoring a link, \ac{BFD} sends lightweight messages in fixed intervals between two switches, connected via a link.
If no packets from the other end of the communication line are received within a defined multiple of the packet \ac{ITT} (i.e. detect multiplier), the link is assumed to have failed.
Here, the \ac{ITT} may be as low as $\SI{1}{ms}$, while the usual detect multiplier amounts to $3$.

Reaction to link failures, discovered by \ac{BFD}, can be realized locally using \ac{OF} \acfp{FFG}.
Therefore, after completing routing of a traffic flow, the controller determines alternative switch configurations for every possible link failure within the main path.
These alternative configurations are stored in the switches' forwarding tables along with the main path using \acp{FFG}.
Thus, in case the outgoing port of a traffic flow is reported as failed, the flow is switched to its alternative path automatically.
To reduce the number of additional forwarding table entries at the switches, our algorithm is designed to maximize the similarity between main and recovery path, letting the traffic flow return to its initial path after as few hops as possible.

\paragraph{SDN-driven Central Recovery}
For centralized link status monitoring, we devise a heartbeat mechanism, similar to \ac{BFD}, which regularly transmits lightweight probe packets.
However, in this case packets are sent out by the controller, thus consuming bandwidth of control and data network.
Encapsulated into \ac{OF}PacketOut messages, heartbeat packets are transferred to the switches, which extract and forward the content on the monitored link.
At the other end of the link, the packet is sent back to the controller using the \ac{OF}PacketIn format.
If this packet is not returned to the controller within a defined interval, the link is classified as failed.

In contrast to local failover, recovery paths are not pre-computed, but determined on-demand, considering current network load for obtaining load/latency optimal routes.

\paragraph{Two-Stage Hybrid Recovery}
Local failover mechanisms usually achieve faster traffic recovery compared to centralized approaches.
Yet, they might employ sub-optimal paths, resulting in network overloads.
Vice versa, controller-driven recovery enables optimal traffic configuration at all times, while failover times are considerably higher.
Subsequently, a hybrid approach presents an intuitive solution, combining the advantages of local and central mechanisms in a divide and conquer manner.  
First, \ac{BFD} is employed for detecting link failures locally.
Hence, traffic can be switched immediately to intact paths with the help of \acp{FFG}.

Next, the controller is notified of the failure.
Subsequently, new globally optimal paths are determined and the switches' forwarding table entries are updated.
Thus, fast recovery is realized, while time intervals of sub-optimal traffic flow, respectively network configuration, are minimized.
To this end, independent fast local protection is combined with globally controlled restoration.

\subsection{Load Optimal Smart Grid Multicast Coordination}
\label{sec:approach:multicast}
Applying multicast flows allows for significant network load reductions.
This is achieved by utilizing a shared path for packets from one source to multiple destinations for as long as prudent.
While this concept is well-known in conventional communication networks, it is applied infrequently due to the significant effort associated with the configuration and management of multicast groups.
However, this technique plays an important role in \ac{IEC} 61850-based communication, being applied for the distribution of measurement values and status updates.

In this work, setup and maintenance of multicast groups is facilitated by providing direct access via the \ac{SDN} \ac{NBI}, as detailed in Section \ref{sec:approach:nbi}.
The Smart Grid application simply has to provide a list of intended group members in terms of \ac{IP} or \ac{MAC} addresses along with a set of packet matching criteria.
After reception of the first packet, which matches the multicast group, the controller performs routing and forwarding rule setup.
To enable multicast handling, paths are defined as routing trees.
For routing, we implemented the \ac{BSMA} \cite{bsma}, which minimizes the number of used links, while at the same time fulfilling flow requirements such as maximum delay bounds.

\subsection{Network Calculus-Based Delay Supervision and Routing}
\label{sec:approach:nc}
Other than in legacy networks, where \ac{NC} can be applied for offline performance evaluation only, \ac{SDN} allows for utilizing this analytical technique during live operation.
For this purpose, we integrate \ac{NC} logic into the \ac{SDN} controller to achieve -- guaranteed -- compliance to defined real-time requirements of Smart Grids at all times. 
A corresponding overview of latency demands is given in Table \ref{tab:timereq} with requirements ranging from \SI{3}{ms} to more than \SI{1}{s}.
To pursue the goal of real-time capable communications, \ac{NC} is applied for the following two use cases:
\begin{itemize}
	\item \textbf{routing} of \textbf{new traffic flows}: provide delay-optimal paths, complying with given latency requirements
	\item \textbf{monitoring} of \textbf{existing traffic flows}: ensure delay bound compliance, even when (other) flows are reconfigured or new flows are added
\end{itemize}
Before going into the details of these tasks, necessary extensions and modifications of \ac{NC} are described in the following section.

\subsubsection{Queue Rate and Cross Traffic Extensions to Network Calculus}
Complex Smart Grid infrastructures and diverse traffic flows require a detailed study of cross traffic impact as they may lead to non-feed forward behavior \cite{yang2017analyzing}, which continues to be an issue of \ac{NC} analysis \cite{fidler_nc,bouillard_feedforward}.
In addition, the influence of \ac{HTB} scheduling has to be considered in \ac{NC} evaluations.

\begin{figure}[t]
	\centering
	\includegraphics[width=\columnwidth]{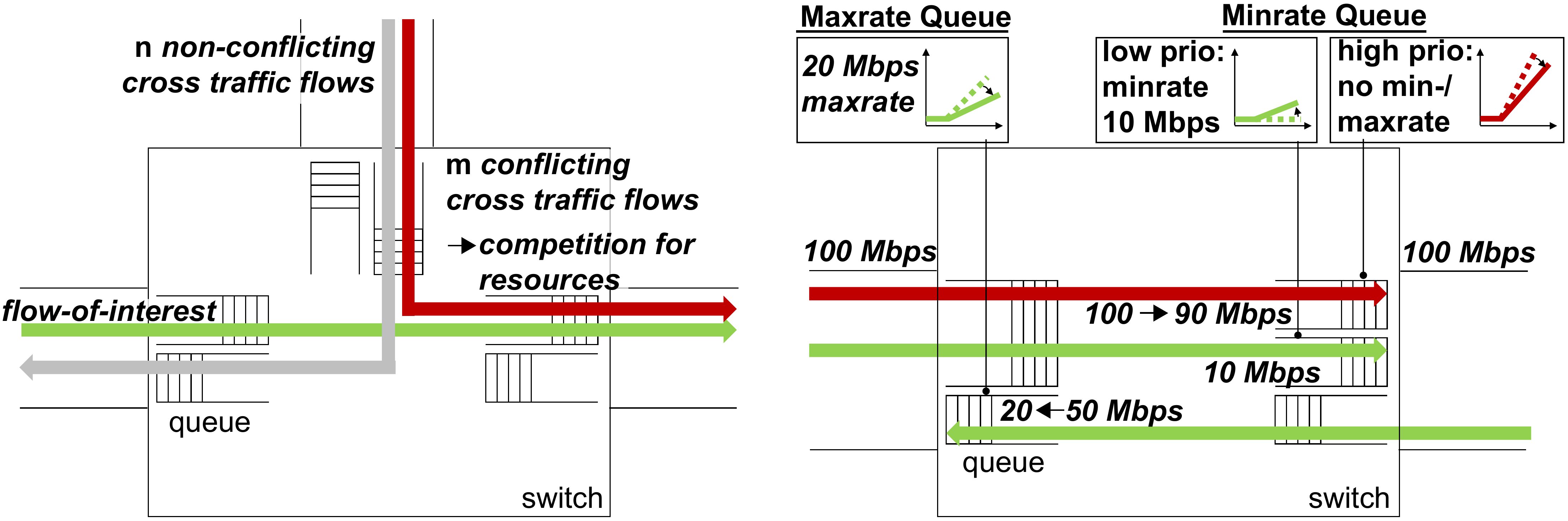}
	\caption{Extensions to Network Calculus: cross traffic handling and integration of queuing with minimum / maximum data rates}
	\label{fig:nc_adv} 
\end{figure}

Beginning with the latter aspect, we enhance our \ac{NC} framework to consider minimum and maximum queue rates as introduced at the beginning of Section \ref{sec:approach}.
Thus, preconditions of our testing environment are reflected.
Maximum queue rates limit the sustained data rate of a flow's service curve.
In contrast, minimum queue rates enhance the service available to a flow by reducing the service curves of higher priority flows, as shown by the right side of Figure \ref{fig:nc_adv}.
Equation \ref{eq:queuerates} formalizes this concept for the service curve $\beta_{k,\textit{foi}}$ of a flow-of-interest \textit{foi},

\begin{equation}
\label{eq:queuerates}
\begin{split}
\beta_{k,\textit{foi}}(t)=\beta_{k}(t)-\sum^{\forall q | p_q\geq p_{\textit{foi}}}\left(\min\left(\sum^{\forall i\in q }\alpha_{k,i}(t),\alpha_{maxDR,q}\right)\right)\\
-\sum^{\forall q | p_q<p_{\textit{foi}}\cap \exists minDR_q}
\left(\min\left(\sum^{\forall i\in q }\alpha_{k,i}(t),\alpha_{minDR,q}\right)\right),
\end{split}
\end{equation}

with $\beta_k$ being the basis service curve at node $k$.
The service available to the flow-of-interest is reduced by the impact of traffic in queues $q$ with same or higher priority ($p_q\geq p_{\textit{foi}}$), considering the sum of respective arrival curves $\alpha_{k,i}$ of flows $i$.
Yet, this influence may be limited by maximum queue rates $\alpha_{maxDR,q}$.
Additionally, flows of lower priority ($p_q<p_{\textit{foi}}$) can curtail the service by up to the corresponding minimum queue rate $\alpha_{minDR,q}$. 

To enable the analysis of non-feed forward networks, we enhance our modeling approach as illustrated by the left side of Figure \ref{fig:nc_adv}.
In \ac{NC} such systems can be assessed with the help of specialized approaches only (e.g. time stopping method), as recursive calculation of cross traffic output curves may lead to deadlocks \cite{fidler_nc}.
Here, this issue is avoided by considering only those cross traffic flows, which use the same output port as the flow-of-interest.
We base this modification on the assumption that interference from other traffic flows at the switches' processing unit is negligibly small.
This hypothesis is confirmed experimentally -- for our testing environment -- by the evaluations in Section \ref{sec:eval:nc:pre}.
In this way, analysis of cross-traffic in non-feed forward networks is converted back into a feed-forward problem.
The associated definition of the left-over service curve $\beta_{k,\textit{foi}}$ for the flow-of-interest \textit{foi} at node $k$ is given by Equation \ref{eq:crosstraffic},

\begin{equation}
\label{eq:crosstraffic}
\beta_{k,\textit{foi}}(t)=\beta_{k}(t)-\sum^{\forall i | k_i+1=k_{\textit{foi}}+1}\alpha_{k,i}(t),
\end{equation}

where the node's basic service curve $\beta_k$ is reduced by the arrival curves $\alpha_{k,i}$ of cross traffic, which shares the same subsequent node $k_i+1$ as the flow-of-interest.

\subsubsection{Traffic and Network Modeling}
\label{sec:approach:nc:model}
As described in Section \ref{sec:ncprincples}, arrival and service curves are modeled by token bucket and rate latency representations respectively.
To parametrize these curves, preliminary measurements are performed, obtaining key traffic and data processing characteristics.
Service is assessed for single traffic flows as well as under full load, as shown in Section \ref{sec:eval:nc:pre}.
In addition, the \ac{SDN} controller performs continuous measurements, verifying the present modeling assumptions.
To this end, \ac{OF} functionalities for collecting port and flow statistics are applied.
Also, information from \ac{LLDP} packets, utilized for topology discovery and updates, is considered.
Finally, heartbeat packets from centralized fast failure detection can be put to use as well.
Thus, compliance of \ac{NC} modeling with the actual network and traffic performance is validated in real-time.
If necessary, the controller may modify service curve parameters to adapt to changed network conditions.
However, adjustments are restricted by measurement cycles and may not be sufficiently fast in case of sudden changes.

\subsubsection{Network Calculus Application in the \ac{SDN} Controller}

\begin{figure}[t]
	\centering
	\includegraphics[width=.7\columnwidth]{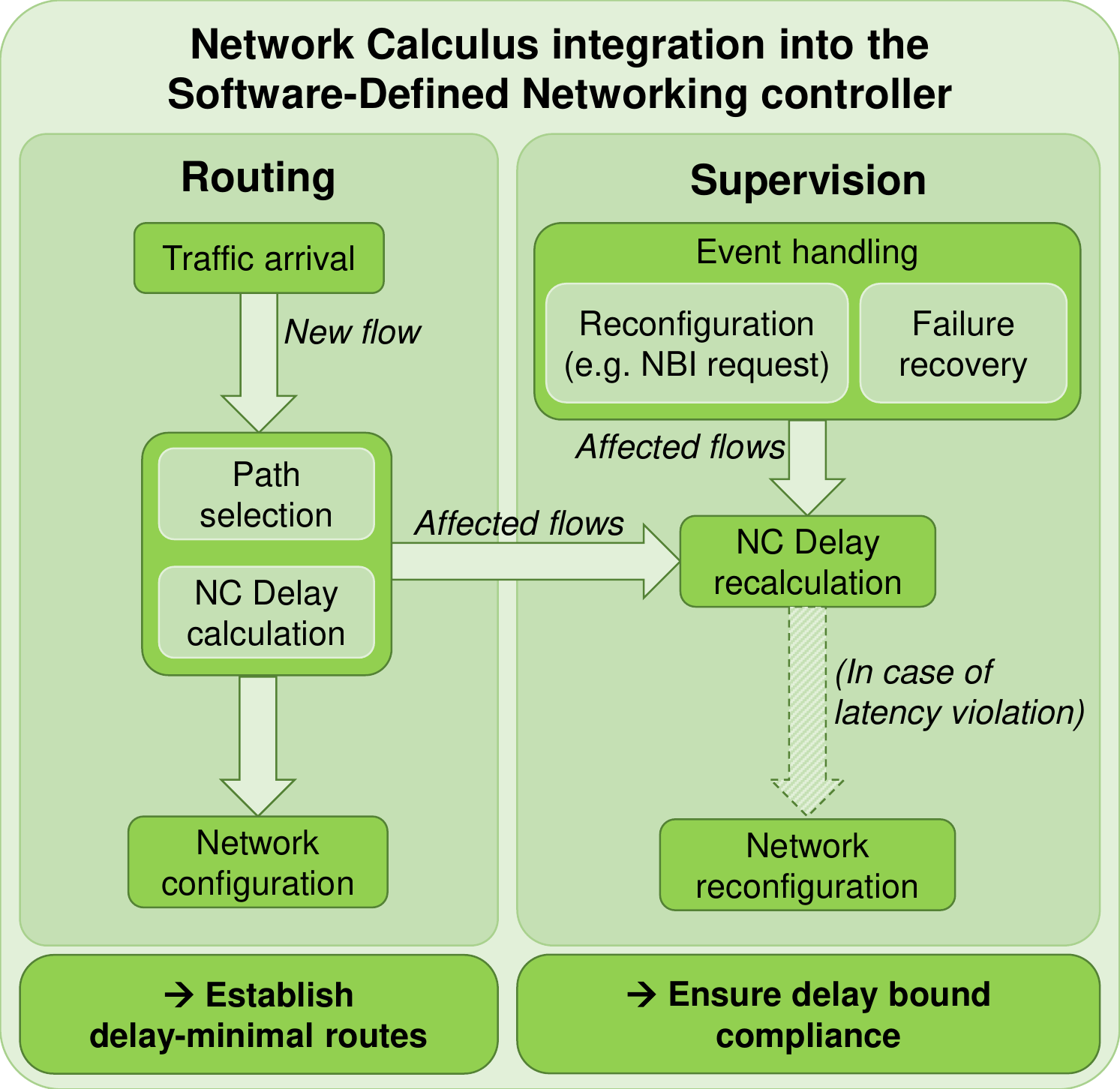}
	\caption{Concept for \acl{NC} integration into the \ac{SDN} controller}
	\label{fig:sdn_nc_int} 
\end{figure}

Figure \ref{fig:sdn_nc_int} gives an overview of the aims and different steps of \ac{NC} integration.
On the arrival of a new traffic flow, the \ac{SDN} controller applies \ac{NC}-based routing to select a
delay-bound compliant path.
We distinguish two different approaches for this task.
Using the concept of full \ac{NC} routing, the new flow's \ac{NC} delay bounds are determined for every path provided by the \ac{DFS}.
Subsequently, the path with the lowest \ac{NC} delay bound is chosen. 
In contrast, the hybrid \ac{NC} routing approach couples standard service-aware routing and \ac{NC} analysis.
In this way, the delay-optimal path is selected by standard routing.
Subsequently, the corresponding \ac{NC} delay bound is calculated for this path only.
If \ac{NC} analysis does not indicate a potential violation of the given latency requirement, the selected route is configured in the network.
Vice versa, if \ac{NC} analysis does indicate a violation, the next optimal path, provided by service aware routing, is assessed.
However, this step incurs additional delay in the routing process.
Eventually, if \ac{NC} routing is not able to find a suitable path for the flow, it would be dropped.
It has to be emphasized that this case would apply to low priority flows only, as paths chosen for high priority flows would be cleared.

Meanwhile, cross traffic, affected by the new flow, is handed over to \ac{NC} delay supervision.
In addition, delay supervision handles flows affected by network reconfiguration.
This applies for example in case of \ac{NBI}-induced modified flow priorities or failure recovery.
In all of the above cases, \ac{NC} delay bounds of affected traffic are recalculated.
If given latency requirements are exceeded, network reconfiguration is triggered.
This involves measures such as rerouting and change of queues (priorities).

For both, routing and delay supervision, performance can be enhanced by re-using previously calculated output bounds of cross traffic flows.
Thus, calculations are sped up, whereas the recalculation of output bounds is not time critical and can be scheduled for subsequent execution.
Detailed performance comparisons of the different routing approaches are provided in Section \ref{sec:eval:nc:rout}.

\begin{algorithm}
	\caption{Network Calculus Delay Supervision Algorithm}
	\label{alg:nccheck}
	\KwIn{Flow f, path p}
	\KwResult{NC delay bound}
		$fPrio \leftarrow getPriority(f)$\\
		\For{$l$ in $getLinksInPath(p)$}{
			\For{$cT$ in $crossTraffic$}{
				\If{outputCurves.contains(cT)}{
					$cToC \leftarrow getOutput(cT)$
				}
				\Else{
					$cToC \leftarrow computeOutputRecursive(cT)$
				}
				\If{$getPrio(cT)>fPrio$}{
					$cToC \leftarrow boundByMaxRate(cToC)$\\
					$highLowPrio \leftarrow add(highLowPrio,cToC)$
				}
				\ElseIf{$getPrio(cT)<fPrio$}{
					$cToC \leftarrow boundByMin(cToC)$\\
					$highLowPrio \leftarrow add(highLowPrio,cToC)$
				}
				\Else{
					$samePrio \leftarrow add(samePrio,cToC)$
				}
				$markForRecalculation(cT,l)$	
			}
			$serviceCurve \leftarrow getServiceCurve(f,l)$\\
			$leftoverSC \leftarrow serviceCurve-highLowPrio$\\
			$leftoverSC \leftarrow getFIFOService(sc,f,samePrio)$\\
			$scETE \leftarrow convolve(scETE,leftoverSC)$
		}	
		$ac \leftarrow getArrivalCurve(f)$\\
		$delay \leftarrow getDelay(ac,scETE)$\\
		\For{$cT$ in $markedDelayBounds$}{
			\If{$lastLatency(cT)+TH>maxLatency(cT)$}{
				$recalculateDelay(cT)$
			}
		}
		$scheduleRecalculation(markedOutBounds)$
\end{algorithm}

\subsubsection{Delay Analysis Algorithm}
Algorithm \ref{alg:nccheck} provides the main steps of our optimized \ac{NC} delay analysis, which is applied for delay supervision and routing.
The links of the intended path are iterated sequentially and checked for potential cross traffic (lines 2-3).
To reduce computation times, previously computed output curves may be used for modeling cross traffic (lines 4-6).
In case of non-optimized processing, or if the curve has not been determined yet, recursive calculation of cross traffic output bounds is required (lines 7-9).
They are computed up to the point of interference with the flow-of-interest.
Next, cross traffic is classified with regard to its priority relative to the flow-of-interest and, if applicable, the service rate is bounded due to minimum/maximum queue rates (lines 10-20).
Also, cross traffic flows are marked for output/delay bound recalculation as the flow-of-interest influences these flows vice versa (line 21).
Afterwards, the base service curve for the flow-of-interest at the current node is retrieved (line 23).
Cross traffic impact is determined according to Equation~\ref{eq:crosstraffic}, using the corresponding output curves with respect to their relative priority (lines 24-25).
By convolving individual service curves the end-to-end bound is calculated (Equation \ref{eq:nc_etesc}).
The arrival curve, in conjunction with the end-to-end service curve, serves as input for deducing the flow-of-interest's upper delay bound (lines 28-29).
Finally, delay bounds of critical flows, which are effected by the flow-of-interest, are recalculated (lines 30-34) and output bound recalculation is scheduled (line 35).

Overall, \ac{NC} allows for predicting and avoiding potential violations of delay bound guarantees, whereas network operation based on measurements reacts to arisen issues only.
Also, measurements provide a snapshot view of the system.
This might be misleading if flows show volatile behavior and measurement intervals are not sufficiently small.
In contrast, increased sampling rates lead to high traffic load on the control network \cite{analysis_vs_measure}.

\section{Smart Grid Reference Scenario and\\Mapping on a Corresponding Communication Infrastructure}
\label{sec:scenario}

\begin{figure}[b!]
	\centering
	\includegraphics[width=\columnwidth]{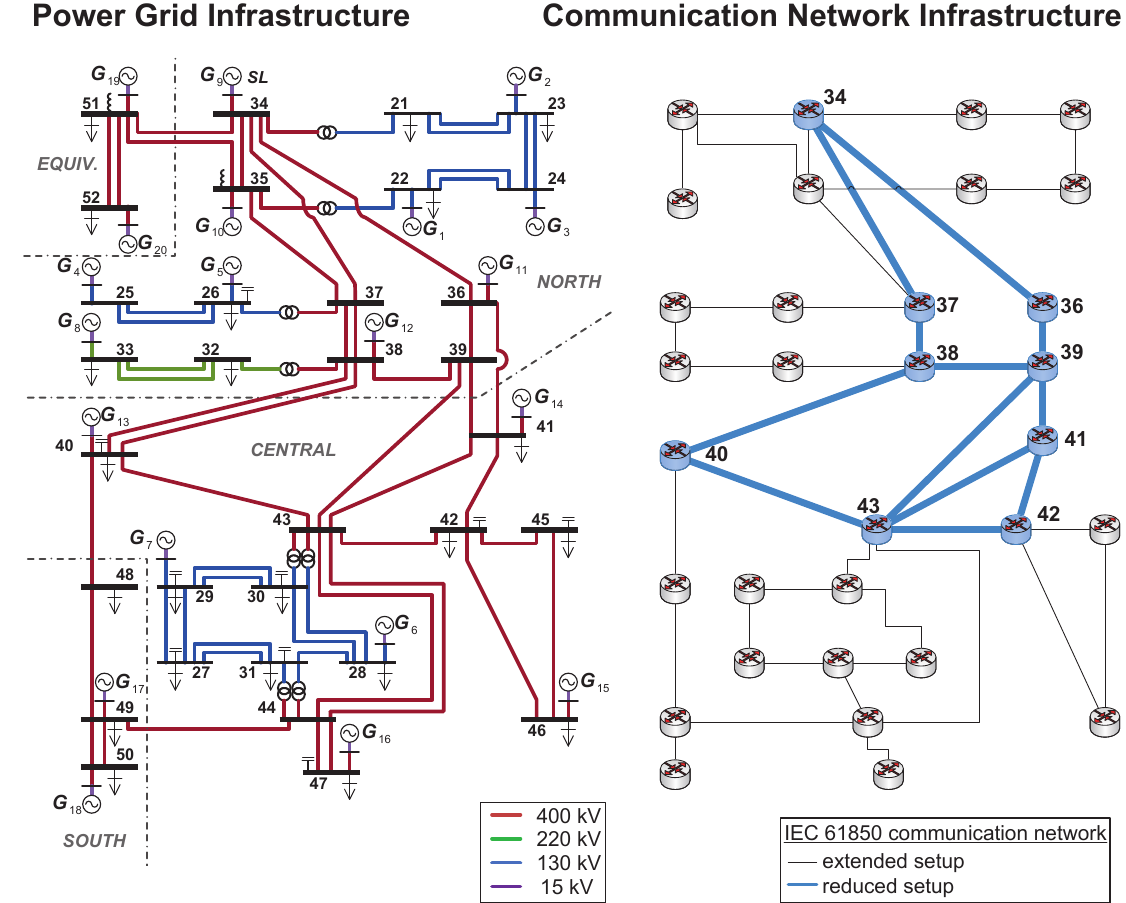}
	\caption{Mapping of the Nordic 32 Test System \cite{nordic32} for power grids to a corresponding IEC~61850-based ICT infrastructure}
	\label{fig:nordic32merged} 
\end{figure}

\paragraph{Topology}
For evaluation we use the Nordic 32 test system \cite{nordic32}, shown on the left side of Figure \ref{fig:nordic32merged}.
The system, derived from actual Swedish and Nordic systems, is well-established for power grid analysis.
It spans four voltage layers from \SI{400}{kV} (red lines) to \SI{15}{kV} (purple lines).
The system is characterized by long \SI{400}{kV} transmission lines and utilizes a nominal frequency of \SI{50}{Hz}. 
Though the test system was originally specified in 1995, it remains valid as the underlying topology is not impacted directly by recent developments towards Smart Grids.
Since it maps higher voltage levels, integration of \acp{DER} is considered in terms of adjusted distribution system loads.
As shown by several current publications, the Nordic 32 test system is still very relevant for power system analysis today \cite{nordic32_cutsem,nordic32_robitzky,nordic_ospina,nordic_peric}. 
Further, it is supported by the fact that the Nordic 32 test system is part of the \ac{IEEE} \ac{PES} 2015 technical report on \textit{Test Systems for Voltage Stability Analysis and Security Assessment} \cite{nordic32_techrep}. 
This lasting relevance of power grid test systems may be attributed to significantly longer innovation cycles \cite{30_years} compared to the \ac{ICT} sector.
Additional details on the specifics of the system can be found in \cite{nordic32_techrep}.

\begin{figure}[b!]
	\centering
	\includegraphics[width=\columnwidth]{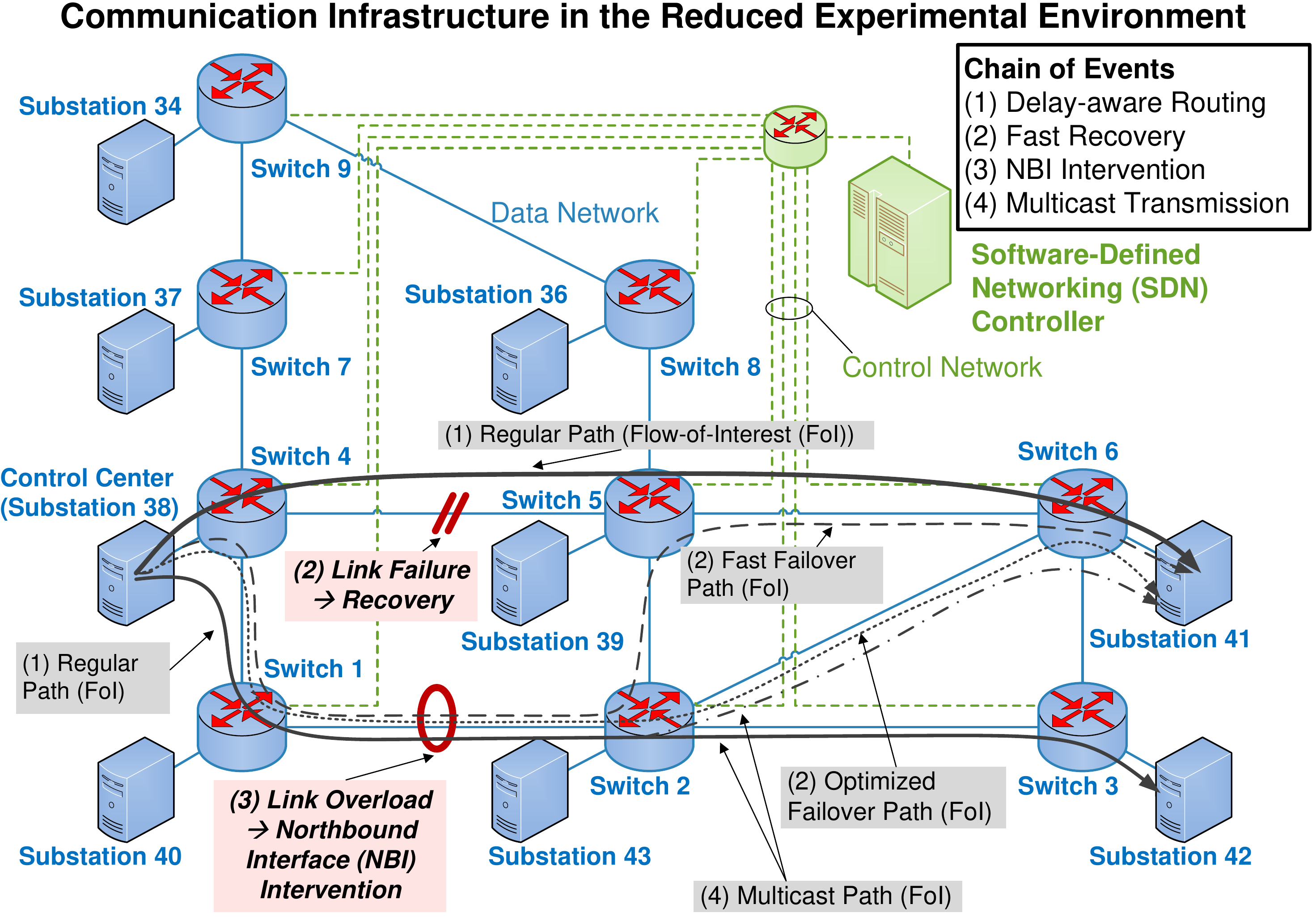}
	\caption{Reduced experimental realization of the communication network's data and control plane including use case specific paths of a flow-of-interest}
	\label{fig:nordic32testbed} 
\end{figure}

On top of this power system, we map a corresponding wide-area communication network infrastructure, shown on the right side of Figure~\ref{fig:nordic32merged}. Networking devices are placed at each substation and connected using fiber-optic cables, carried along the power lines.
Thick, blue lines highlight an excerpt of the network, which is modeled in our empirical testbed setup, using dedicating hardware for each network device.
Scaling of the scenario to the entire network (grey lines) is achieved 1) in the testbed setup by running two virtual switches on the same server hardware and 2) by utilizing Mininet emulation \cite{mininet}, where applicable.
Figure \ref{fig:nordic32testbed} details the small-scale testbed implementation, while Section \ref{sec:testbed} provides hardware specifications.
 
\paragraph{Traffic Pattern}
Traffic patterns (number of flows, communication partners) for this evaluation scenario are generated on basis of relevant, real-world transmission grid functionalities \cite{smartgrid_traffic,Budka_smartgrid_traffic}.
Several of these applications are already in use in today's power grids, whereas others are regarded viable for deployment in future Smart Grids.
In all cases, standard protocols are considered.

\begin{table}[b!]
	\centering
	\caption{Traffic patterns for Nordic 32 test system}
	\begin{tabular}{c c c c c }
		\toprule
		\parbox{1.6cm}{\centering Message Type} & Source(s) & Destination(s) & \parbox{1.9cm}{\centering Number of flows in reduced (extended) experiment} & \parbox{2cm}{\centering Scenarios (Sections)} \\
		\midrule
		GOOSE & 38 & all & 8 (31) & 1-4 (Sec. \ref{sec:eval:failover}-\ref{sec:eval:nc})\\
		\midrule
		SV & all & 38 & 8 (31) & 1-4 (Sec. \ref{sec:eval:failover}-\ref{sec:eval:nc})\\
		SV & all & neighbors & 23 (85) & 1-4 (Sec. \ref{sec:eval:failover}-\ref{sec:eval:nc})\\
		\midrule
		MMS & 38 & 34, 42 & 2 (8) & 2 (Sec. \ref{sec:eval:nbi})\\
		\midrule
		MAS & 38 & 41, 42, 43 & 3 (3) & 2-4 (Sec. \ref{sec:eval:nbi}-\ref{sec:eval:nc})\\
		MAS & 39 & 34, 36, 43 & 3 (3) & 2-4 (Sec. \ref{sec:eval:nbi}-\ref{sec:eval:nc})\\	
		MAS & \multicolumn{2}{c}{(further MAS groups)} & (17) & 2-4 (Sec. \ref{sec:eval:nbi}-\ref{sec:eval:nc})\\	
		\midrule
		\textbf{Total} & & & 47 (178) & \\
		\bottomrule	
	\end{tabular}
	\label{tab:trafficpat}
\end{table}

\ac{SCADA} incurs communication from the control center to every substation and vice versa to obtain measurement data and perform remote control \cite{scada}.  
Here, we utilize \ac{IEC}~61850 communication services for this purpose, as suggested in \cite{61850-90-2:TC57:IEC}.
In particular, control commands from the control center, situated at Substation 38, are sent to all substations using \ac{GOOSE} messages.
\ac{SV} serve for exchanging measurement data with the control center as well as between neighboring substations.
The latter is required for inter-substation protection functions, such as current differential protection \cite{diff_protect}.
Starting from Subsection \ref{sec:eval:nbi}, \ac{MAS} messaging is introduced for distributed power flow control within multiple clusters of substations \cite{robitzky_agent}.
Also, \ac{MMS} transmissions are considered for configuration and software update purposes.
Though there may be additional traffic, e.g. enterprise voice and data communications, we limit our analysis to the critical functions outlined above.
Table \ref{tab:trafficpat} sums up used traffic patterns.

\paragraph{Sequence of events}
In addition, Figure \ref{fig:nordic32testbed} visualizes the following sequence of use cases, considering \ac{GOOSE} traffic from the control center (Substation 38) to Substation 41 as flow-of-interest for this analysis:
\begin{enumerate}
	\item \textbf{Delay-aware routing} provides the primary path for this flow via Substations 38, 39, 41 (solid lines).
	\item This path is interrupted by a \textbf{failure} between Substations 38 and 39, resulting in \textbf{recovery} to the fast (dashed lines) and the optimized failover path (dotted lines) (Section \ref{sec:eval:failover}).
	\item Evoked by the failure, combined with additional \ac{MAS} and \ac{MMS} traffic, the link between Substations 40 and 43 is \textbf{overloaded}.
	To maintain grid stability, dynamic \textbf{re-configuration} -- triggered via the \textbf{\ac{NBI}} -- needs to be carried out (Section \ref{sec:eval:nbi}).
	\item Finally, dash-dotted lines illustrate load optimization on basis of \textbf{multicast} transmission (Section \ref{sec:eval:multicast}). 
\end{enumerate}

\section{Evaluation Environment for Empirical Performance Assessment}
\label{sec:testbed}
This section sums up the most important characteristics of our experimental environment as well as the used emulation software.
Each experiment respectively emulation is repeated 100 times with a duration of \SI{60}{s}, typically resulting in up to 6 million data points per traffic flow. 

\subsection{Experimental Set-up}
Our experimental environment, shown in Figure \ref{fig:nordic32testbed}, consists of three independent networks: data, control and management, created in hardware.
The first network covers the data plane of the \ac{SDN} architecture, representing the wide-area infrastructure for transmitting Smart Grid traffic.
It includes up to 28 \acp{vSwitch}, running \acf{OVS} v2.5.2 under Ubuntu 16.04.2 LTS (v4.4.0-77-generic x86-64 Kernel).
The \acp{vSwitch} are deployed on 14 servers with standard hardware (Intel Xeon D-1518 with one two port I210-LM and two four port I350 Intel 1GBase-T Ethernet \acp{NIC}).

\begin{figure}[t]
	\centering
	\includegraphics[width=.7\columnwidth]{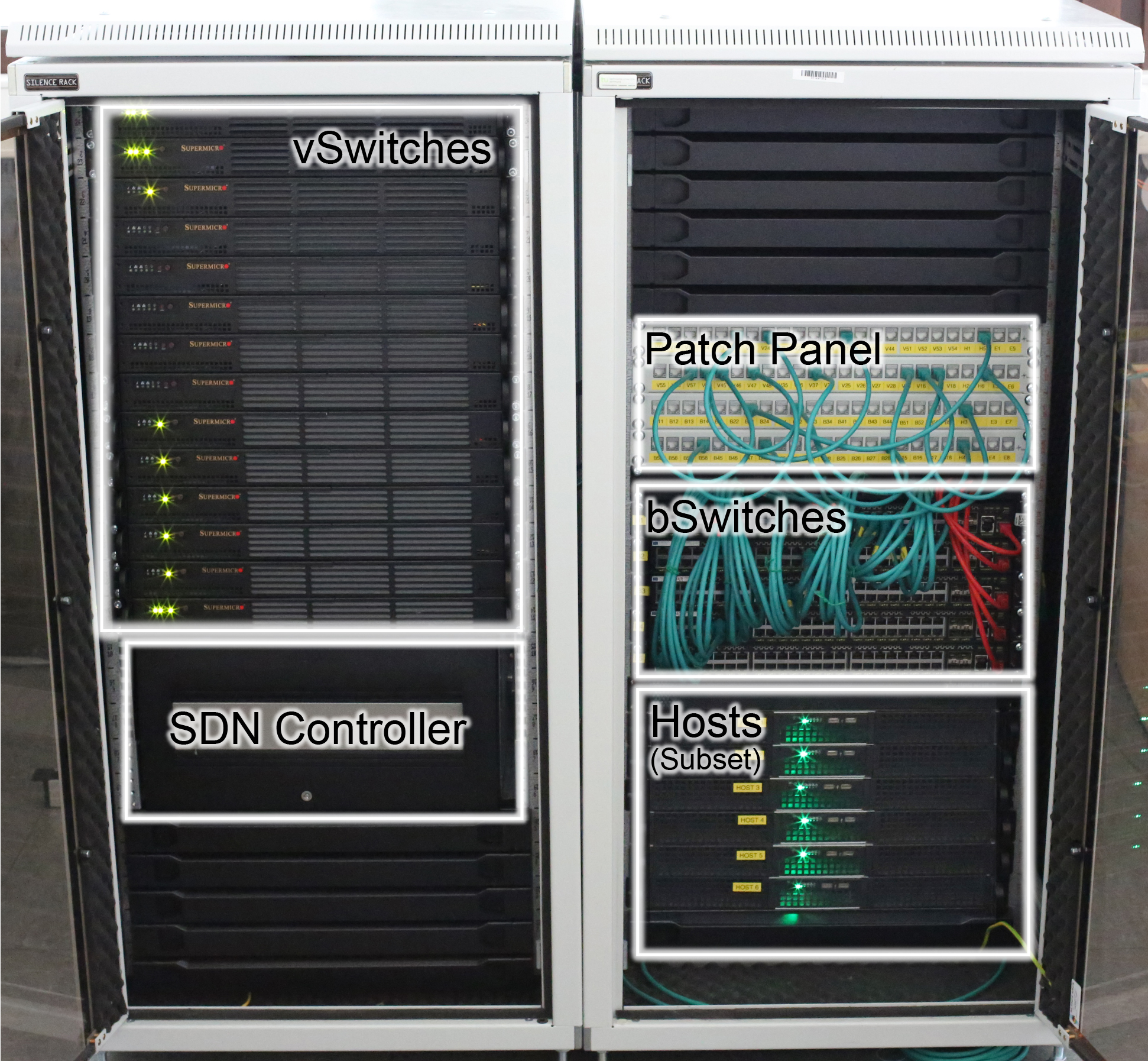}
	\caption{Experimental testing environment for SDN in Smart Grids}
	\label{fig:testbed} 
\end{figure}

The reduced set-up is limited to four \acp{vSwitch}, each run on an individual server.
In comparison, for the extended environment one server is required to host two switches simultaneously.
In this case, every \ac{vSwitch} is assigned exclusive ports on separate \acp{NIC} as well as dedicated \ac{CPU} cores.
Thereby, effective isolation of network hardware is ensured.
According to \cite{vm_overhead} virtualization overheads can be classified negligible for the purposes of this work. 
In addition, we deploy five 48 port Pica8 3290 \acp{bSwitch}, which utilize \ac{OVS} v2.3.0 under PicOS 2.6.32.
The data network is completed by seven dedicated hosts, six of which are Intel Celeron J1900 with a two port I210-LM \ac{NIC}.
To achieve timing precision in the range of a few microseconds, while avoiding synchronization issues, the seventh host (Intel Xeon D-1518) models Substations 38 and 41 simultaneously.
Thus, corresponding measurements utilize a single clock.
For mapping the entire Nordic 32 system, we additionally employed virtualized hosts on 12 servers (Intel Xeon X5650).

The \ac{SDN} control plane is constituted by an out-of-band network and a server (Intel Xeon D-1518), hosting the \acs{SUCCESS} platform.
Connection to the switches of the data plane is established using OpenFlow v1.3. 

Finally, the management network enables remote configuration, starting and stopping of measurement processes at all hosts.
Hence, it is applied for facilitating the experiment and is not part of the evaluations itself.
For both, the control and the management network, one Zyxel GS1900-24E switch each provides Gigabit connectivity.
Abstracting from real-world scenarios, copper instead of fiber-optic cables are employed.
An overview of our testing environment is given in Figure \ref{fig:testbed} in terms of a photo of the actual laboratory set-up.

\subsection{Network Emulation}
To validate the experimental results and conveniently scale certain aspects of evaluation (e.g. control plane performance) to the full Nordic 32 test system, network emulations are carried out.
Therefore, the software Mininet \cite{mininet} is run on an Intel Xeon D-1518 under Ubuntu 16.04.2 LTS (v4.4.0-77-generic x86-64 Kernel).
Mininet allows for the set-up of complex, realistic network configurations, applying the same controller framework as in the experiment.
Configuration is performed using the Python programming language.

\section{Evaluation of Approaches Proposed for\\Mission Critical Communications}
\label{sec:results}
Evaluation is split into four parts, each highlighting different hard service guarantee aspects, introduced in Section \ref{sec:approach}.

\subsection{Comparison of Fast Failover Approaches}
\label{sec:eval:failover}

Within this subsection, we compare the failure detection and recovery mechanisms, described in Section \ref{sec:approach}, with regard to recovery delays, route optimality and induced network load.
\acf{BFD} was configured with an \acf{ITT} of \SI{1}{ms} and a detection multiplier of $3$, whereas the controller \ac{HB} does not stabilize until an \ac{ITT} of \SI{3}{ms}, timing out after \SI{15}{ms}.
A link failure between Substations 38 and 39 is produced, interrupting the \ac{GOOSE} traffic flow from the control center to Substation 41.

\begin{figure}[b!]
	\centering
	\includegraphics[width=\columnwidth]{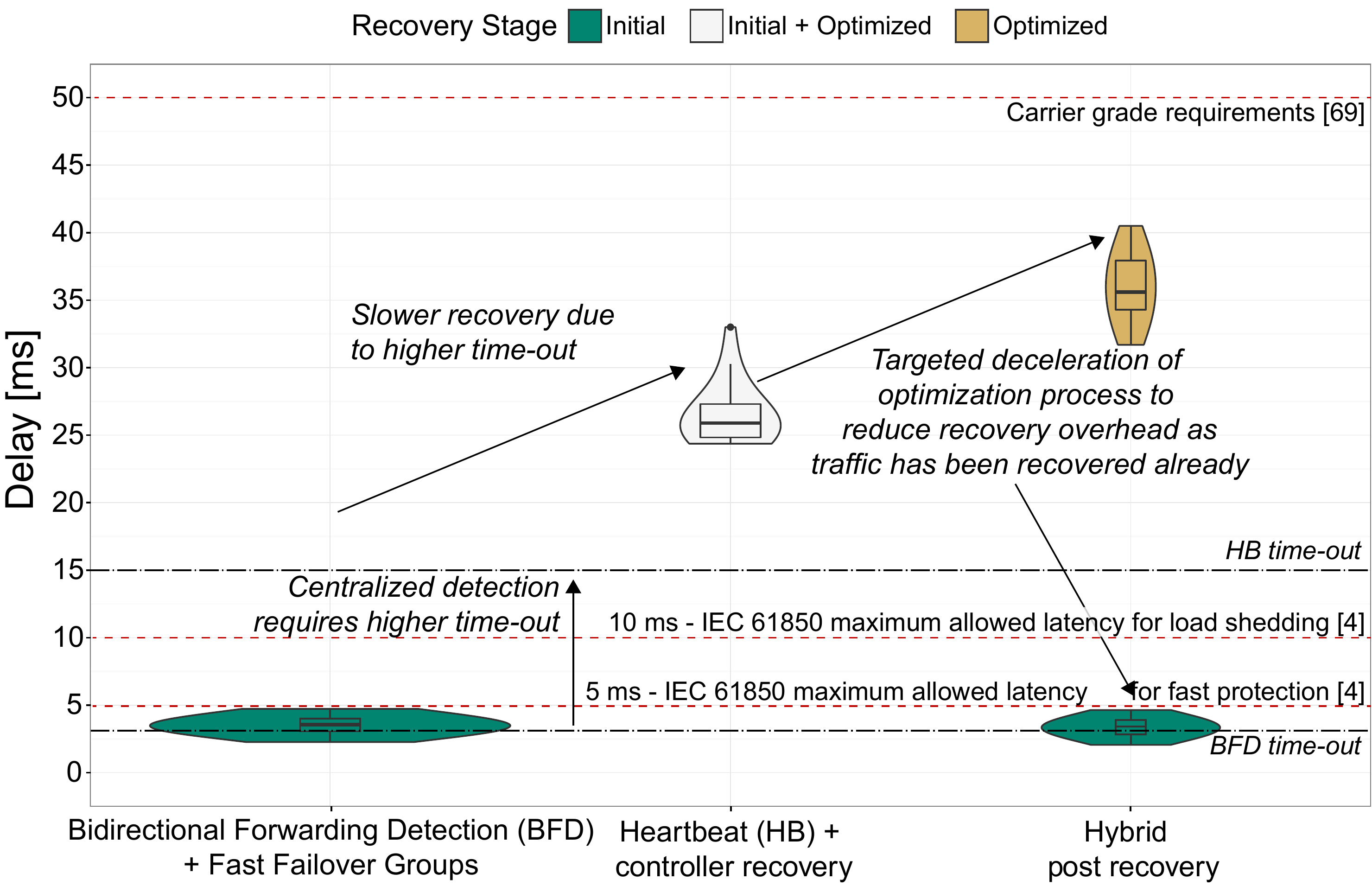}
	\caption{Comparison of initial and optimized recovery delay for different approaches using Software-Defined Networking}
	\label{fig:failover_delay} 
\end{figure}

\subsubsection{Recovery Delay Evaluation}
Figure \ref{fig:failover_delay} depicts the flow's end-to-end recovery delays, measured at Substation 41 in our testbed set-up (c.f. Figure \ref{fig:nordic32testbed}).
End-to-end recovery delay refers to the time difference between the last packet received before the failure and the first packet received after clearance. 
It can be seen that recovery delays depend significantly on the detection mechanism applied.
Using \ac{BFD} traffic is switched to an alternative path within \SI{4.73}{ms} at maximum.

In contrast, controller centric failure detection and recovery requires up to \SI{33}{ms}.
Yet, this approach redirects the \ac{GOOSE} traffic flow to an optimal path directly, whereas applying \acf{FFG} in combination with \ac{BFD} necessitates subsequent optimization.
This step may be triggered in response to the reception of regular OFPortStatus messages, which is not until approximately \SI{350}{ms} after the failure \cite{SDN4SG}.

Integrating the advantages of both approaches, the hybrid approach uses \ac{BFD} and \ac{FFG} for immediate recovery, achieving the same latencies.
In a second step \acf{HB} messages are used to initiate controller-based post optimization with a mean delay of \SI{35.94}{ms}.
This value is close to the recovery delay of the controller centric approach.
To minimize network load of the hybrid approach, the \ac{HB} interval for post optimization is increased to \SI{10}{ms}.
This choice is a trade-off between fast optimization of routes and reduced data and control network load. 
Using this parameter set, optimization is executed within about \SI{40}{ms} at maximum.
Thus, carrier grade requirements (\SI{50}{ms}) \cite{mpls_tp} are fulfilled, while considering a security margin of \SI{10}{ms}.
Faster optimization could be achieved by applying the same values as for controller-driven recovery (c.f. restrictions above).
In contrast, further load reduction could be enabled by increased \acp{ITT} and time-out intervals. 
For example, when striving for the \ac{IEC}~61850 requirement of \SI{100}{ms} for slow automatic interactions, the \ac{ITT} might be raised to \SI{25}{ms} (detect multiplier: 3).
Further details on load reduction are discussed at the end of this subsection.

\subsubsection{Path Optimality}
\begin{figure}[t]
	\centering
	\includegraphics[width=\columnwidth]{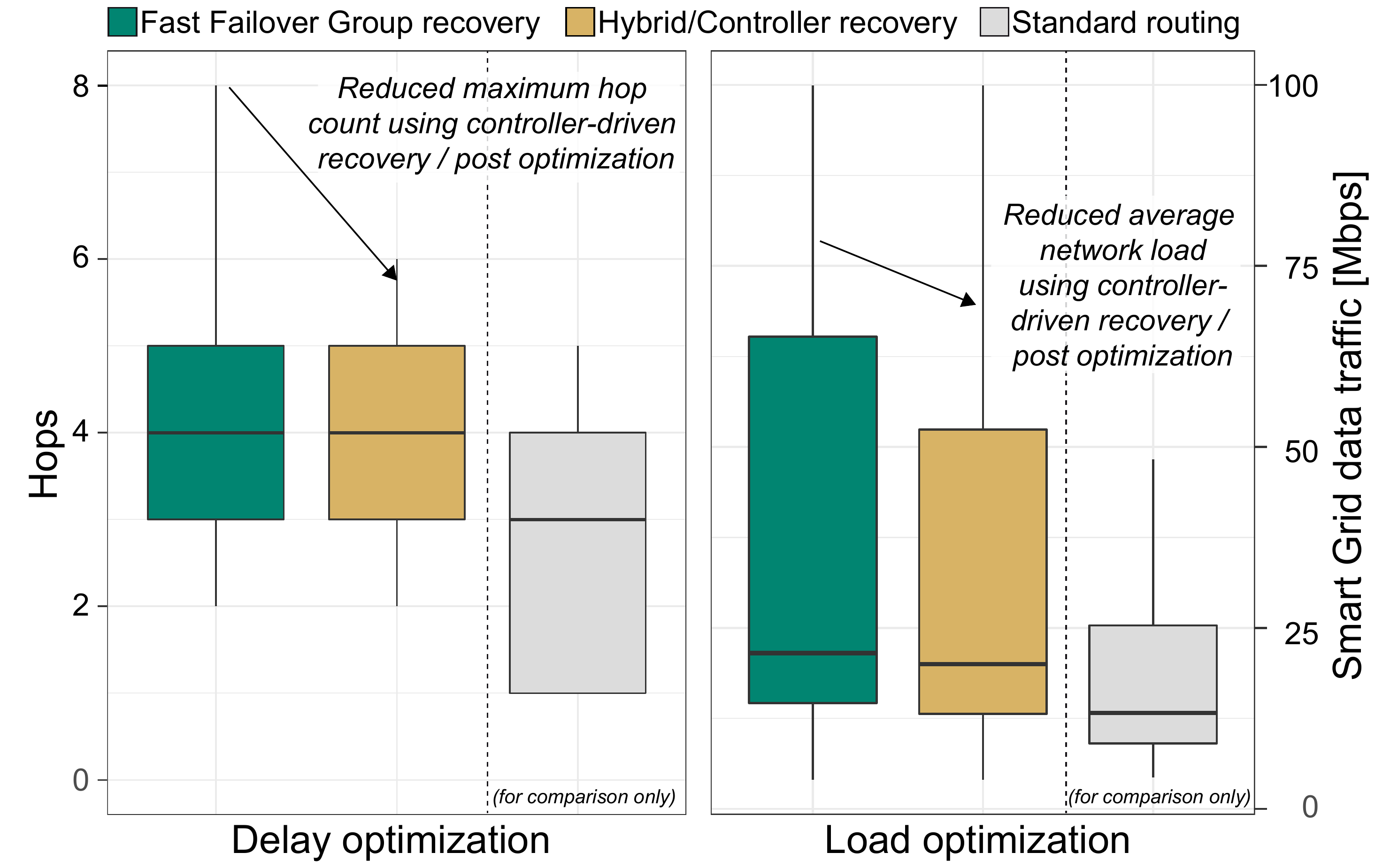}
	\caption{Hop counts and network load before/after failover using different Software-Defined Networking-enabled recovery methods}
	\label{fig:failover_hopcount_load} 
\end{figure}

Figure \ref{fig:failover_hopcount_load} illustrates the aspect of path optimality, considering the criteria minimum hop count (left side) and load balanced network links (right side).
This study utilizes Mininet emulation (hop count), respectively the extended hardware set-up (network load), to study the entire 75 link communication network of the full Nordic 32 system.
The results of regular routing, before the failure, serve as benchmark for both cases.
The left-side of Figure~\ref{fig:failover_hopcount_load} visualizes the increase to a maximum hop count of eight due to \ac{FFG} recovery.
In comparison, the maximum hop count in case of controller recovery amounts to six only.
According to the right side of Figure \ref{fig:failover_hopcount_load}, the median network load is reduced from \SI{22}{Mbps} in case of \ac{FFG} paths, to \SI{20}{Mbps} after controller recovery respectively post optimization.
This effect is highlighted even more clearly by reduced upper and lower quartiles. 

\subsubsection{Link Load Assessment}
Table \ref{tab:ff_addload} sums up the additional network load induced by the different failure detection mechanisms.
Link respectively network utilizations $\eta$ are determined analytically for the monitoring of the entire 75 link Nordic 32 system, using the straightforward approach given in Equation \ref{eq:nl_hb}:

\begin{equation}
\label{eq:nl_hb}
\eta=\dfrac{p}{ITT}\cdot \dfrac{n}{R}
\end{equation}

where for the control network the maximum of OFPacketIn and OFPacketOut message (encapsulating the heartbeat message) is used as packet size $p$.
Raw Ethernet packet size of \ac{BFD}/\ac{HB} messages is applied for the data link. 
$R$ refers to the offered network capacity and $n$ indicates the number of monitored links in case of the control network load. 
For the data network, each link is considered individually, resulting in $n=1$.

\begin{table}[t]
	\centering
	\caption{Continuous additional load due to failure detection mechanisms on 75 data network links with \SI{1}{Gbps} capacity each and \SI{1}{Gbps} control network}
	\begin{tabular}{p{2.3cm} c c c c c c}
		\toprule
		\textbf{Recovery Approach} & \multicolumn{3}{c}{\multirow{2}{*}{\textbf{Data Network}}} & \multicolumn{3}{c}{\multirow{2}{*}{\textbf{Control Network}}} \\
		\midrule
		& \parbox{1cm}{\centering ITT [ms]} & \parbox{1cm}{\centering Packet Size [Bit]} & \parbox{1cm}{\centering Load [\%]} & \parbox{1cm}{\centering ITT [ms]} & \parbox{1cm}{\centering Packet Size [Bit]} & \parbox{1cm}{\centering Load [\%]}\\
		\midrule
		\ac{BFD} & 1 & 560 & 0.056 & - & - & 0 \\
		\midrule
		\parbox{2.3cm}{Controller-Heartbeat} & 3 & 512 & 0.017 & 3 & 1,344 & 3.360 \\
		\midrule
		Hybrid & 1/10 & 560/512 & 0.061 & 10 & 1,344 & 1.008 \\
		\midrule
		\parbox{2.3cm}{Hybrid\\optimized} & 1 & 560 & 0.056 & - & - & 0 \\
		\bottomrule
	\end{tabular}
	\label{tab:ff_addload}
\end{table}

While the controller \ac{HB} achieves the lowest data network load of \SI{0.017}{\%}, its frequent transmissions back to the \ac{SDN} controller require \SI{3.360}{\%} of the control network capacity, which is the highest demand among all approaches.
In comparison, even the hybrid approach, which comprises less frequent \ac{HB} messages, incurs a control network load of just \SI{1.008}{\%}.
However, a slight increase in data network load to \SI{0.061}{\%} has to be noted.
Finally, the data network load of \ac{BFD} is in between the other two approaches, whereas the control network is only stressed in case of failure.
Further optimization of the hybrid mechanism, may reduce its associated network loads to the same levels as those of \ac{BFD}.
Overall, the load on the monitored link is comparatively low in all cases ($<\SI{0.1}{\%}$).
In comparison, the control network could experience considerable stress, depending on its topology and the number of monitored links.
Additionally, assuming adequate processing resources being available to the controller, it needs to be highlighted that scalability of the recovery approaches boils down to the issue of control network utilization.
Corresponding loads are observed to be minor in this work, as a result of applying out-of-band control.
In contrast, it might become a more severe issue, when in-band control is employed in real-world scenarios. 
Hence, in such scenarios hybrid fast failover should utilize reduced \acp{ITT} or the hybrid optimized failover, relying on \ac{BFD} only.

All in all, the hybrid recovery concept can be considered a reasonable compromise between low recovery delays, path optimality and consumed network capacity.
The latter is even improved by an optimized version of the approach. 

%


\subsection{Smart Grid Service-Driven Dynamic Priority Adaption}
\label{sec:eval:nbi}

\begin{figure}[b!]
	\centering
	\includegraphics[width=\columnwidth]{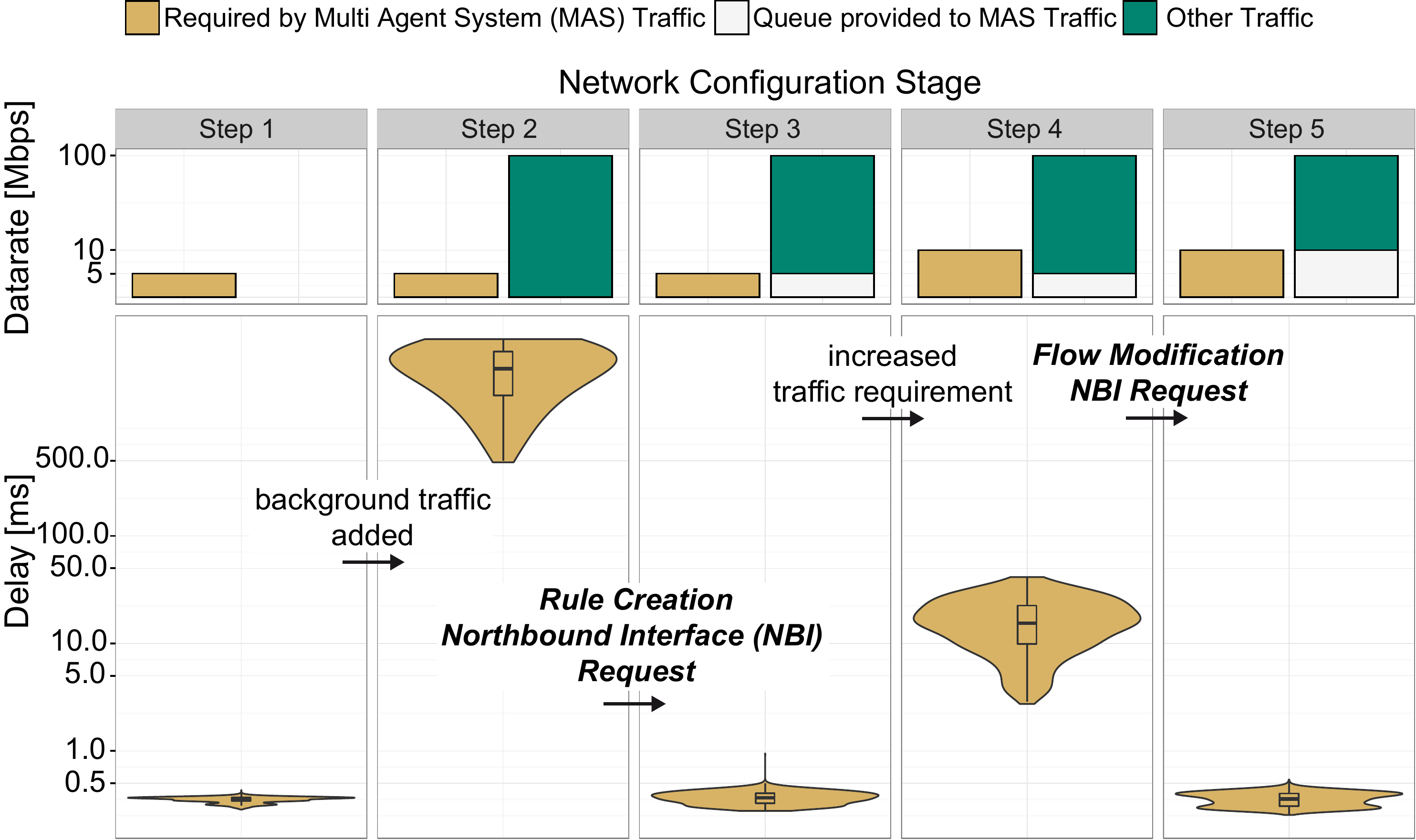}
	\caption{Successive steps of handling \acf{MAS} traffic in response to changing network conditions, \acf{NBI} requests and subsequent priority/queue assignment}
	\label{fig:modifyFlow_dr} 
\end{figure}

Using the example of varying service requirements for \acf{MAS}-based distributed power grid control, dynamic adaption of network configurations is shown.
This involves prioritization, queuing and \acf{NBI} requests.
A five step sequence of dynamic prioritization tasks is executed, as shown in Figure \ref{fig:modifyFlow_dr}.
The sequence involves two of the \ac{NBI} requests, introduced in Section \ref{sec:approach}.

In step 1, \ac{MAS} traffic is transmitted on an empty link between Switches 40 and 43.
In total, these \ac{MAS} messages have a capacity demand of approximately \SI{5}{Mbps}, illustrated by bar plots in the upper part of Figure \ref{fig:modifyFlow_dr}.
This results in mean latencies of \SI{351}{\mu s}, depicted by the violin plots in the graph's lower part.

Next, normal traffic conditions, as described in Section \ref{sec:scenario}, are restored.
Hence, \ac{GOOSE} and \ac{SV} traffic are present on the network as well.
Further, additional \ac{MMS} traffic for the purpose of updating devices is injected into the \ac{ICT} infrastructure.
In conjunction with the link failure, discussed in the previous subsection, this leads to an overload of the communication link between Substations 40 and 43 as shown in step 2 of Figure \ref{fig:modifyFlow_dr}.
Since \ac{MAS} traffic is not recognized by the controller yet, it is handled as best effort, causing a drastic increase of the delay of up to \SI{6.76}{s}.

To resolve this issue, a Rule Creation request is sent.
Thus, the \ac{MAS} priority is raised to $30$, which is well above the priority of \ac{MMS} (priority level 20).
Adequate queues with \SI{5}{Mbps} minimum data rate are arranged for.
Hence, delays are reduced back to below \SI{1}{ms}, as shown in step 3.

Next, due to the power system being highly loaded and not in (N-1) secure state, an outage occurs, disconnecting the transmission line between Substations 38 and 39.
Subsequently, parallel transmission lines between Substation 40 and 43 become overloaded.
This emergency situation is identified by the agents of the distributed control system.
To prevent cascading outages, the \ac{MAS} aims at estimating the grid state on basis of refined measurement data.
Accordingly, its monitoring precision has to be improved.
Building on the detailed view of the power system, adequate counter-measures can be determined, which -- in this case -- involves triggering a \ac{PFC}.
These developments lead to more frequent transmissions of critical \ac{MAS} messages, thus increasing the traffic load, as shown in step 4 of Figure \ref{fig:modifyFlow_dr}.
However, the queue assigned to \ac{MAS} messaging is not sufficient for these altered data rate requirements, causing a rise in delay up to \SI{41.43}{ms}.

Subsequently, a Flow Modification request is issued to obtain a temporary raise of priority.
Thus, \ac{MAS} traffic is switched to a higher priority queue, providing up to \SI{10}{Mbps} minimum data rate and restoring the initial delay level (step 5 of Figure \ref{fig:modifyFlow_dr}).
In this way, despite of the heavily loaded communication network, timely transmission of critical control messages can be ensured.
In turn, power system stability can be maintained, preventing cascading outages.

\subsection{Validation of Multicast Load Reduction}
\label{sec:eval:multicast}

\begin{figure}[t]
	\centering
	\includegraphics[width=\columnwidth]{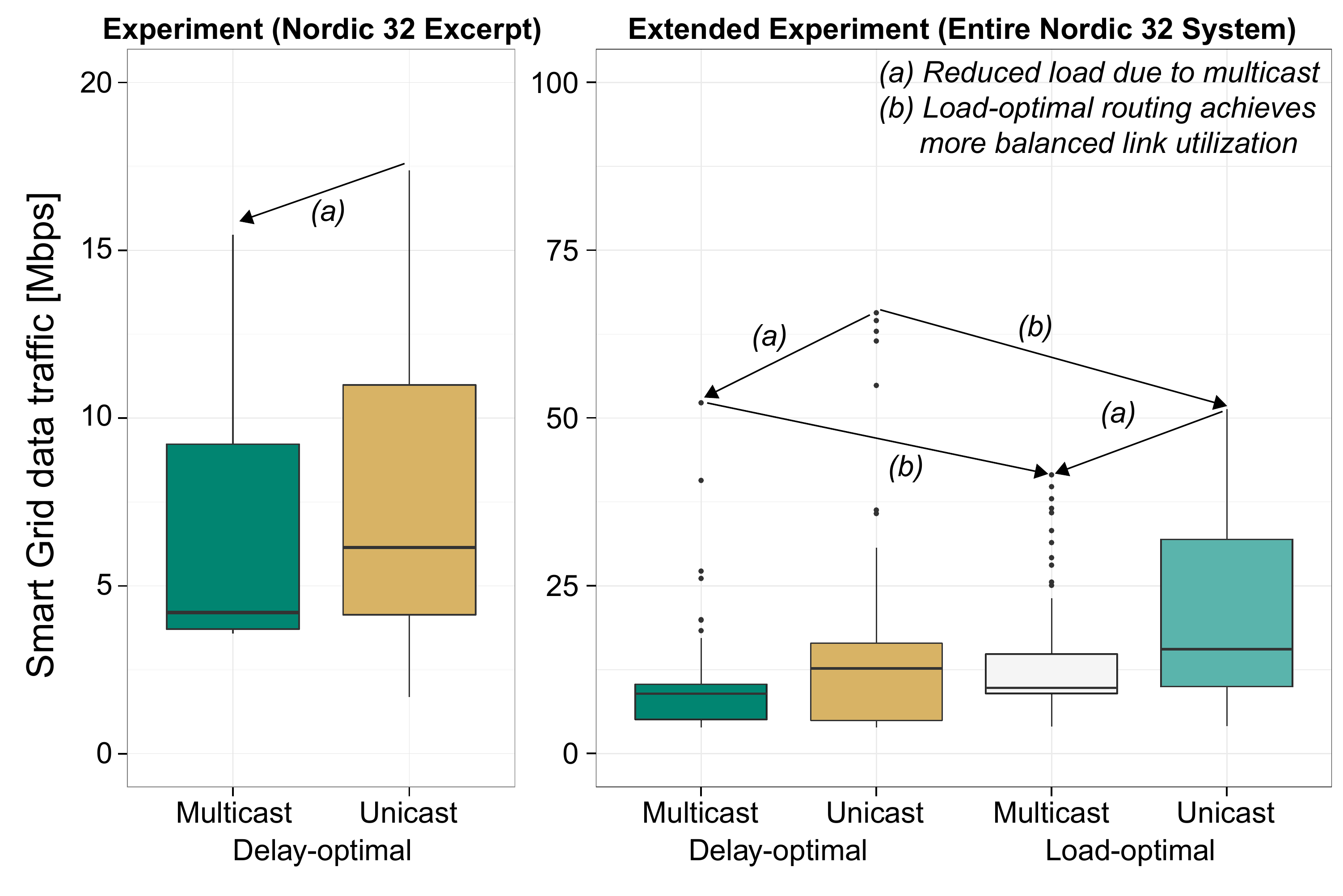}
	\caption{Comparison of network load using uni- and multicast flows in simulation (left) and experiment (right)}
	\label{fig:multicast_load} 
\end{figure}

This subsection targets load reduction with the means of multicast transmission.
Therefore, transfer of measurement values and statuses from one to multiple other substations is bundled in multicast transmissions, wherever possible.
In addition, if identical commands are sent by the control center to several substations, these GOOSE messages are transferred as multicast.
On shared paths between different agents of the distributed control system joint transmission is employed as well.
The resulting optimization of bandwidth consumption in the network is studied using experiments.

Figure \ref{fig:multicast_load} (left side) contrasts network utilization for unicast and multicast transmission, measured in our testbed.
Compared to unicast, the mean link load is reduced from \SI{7.50}{Mbps} to \SI{6.63}{Mbps}.
In addition, applying multicast diminishes the maximum load by \SI{11.1}{\%} to \SI{15.47}{Mbps}, shown by marker (a).

Scaling up, the extended experimental set-up is used to study the impact of multicast on the whole 75 link Nordic 32 system.
While in the reduced testing environment we focus on delay optimal routing, the extended measurements include load optimal routing as well.
Figure \ref{fig:multicast_load} (right side) shows link loads for the four different combinations of uni-/multicast transmission and delay/load optimal routing.
Similar to the previous experiments, a reduction of mean and maximum load is observed, when exchanging unicast for multicast transfers, highlighted by marker (a).
This holds true for both routing disciplines.
Comparing the different routing schemes -- among each pair of unicast respectively multicast transmissions -- shows an increase of mean link utilization for load optimal routing.
In contrast, the maximum load is delimited to a lower level as can be seen from marker (b).
This behavior matches perfectly the concept of balancing network utilization.

\subsection{Evaluation of In-Controller Network Calculus Supervision and Routing}
\label{sec:eval:nc}
As described in Section \ref{sec:approach:nc}, we apply \acf{NC} for delay-aware routing of traffic flows and online supervision of latency requirement compliance.
In the following, prerequisite evaluations are performed for assuring the assumptions of modified cross traffic handling.
Next, calculated \ac{NC} delay bounds are cross-validated against the results of empirical measurements.
This section concludes with evaluations on the applicability and optimization of \ac{NC}-based routing and delay supervision.

\subsubsection{Prerequisite Assessment of Cross Traffic Handling}
\label{sec:eval:nc:pre}

\begin{figure}[t]
	\centering
	\includegraphics[width=\columnwidth]{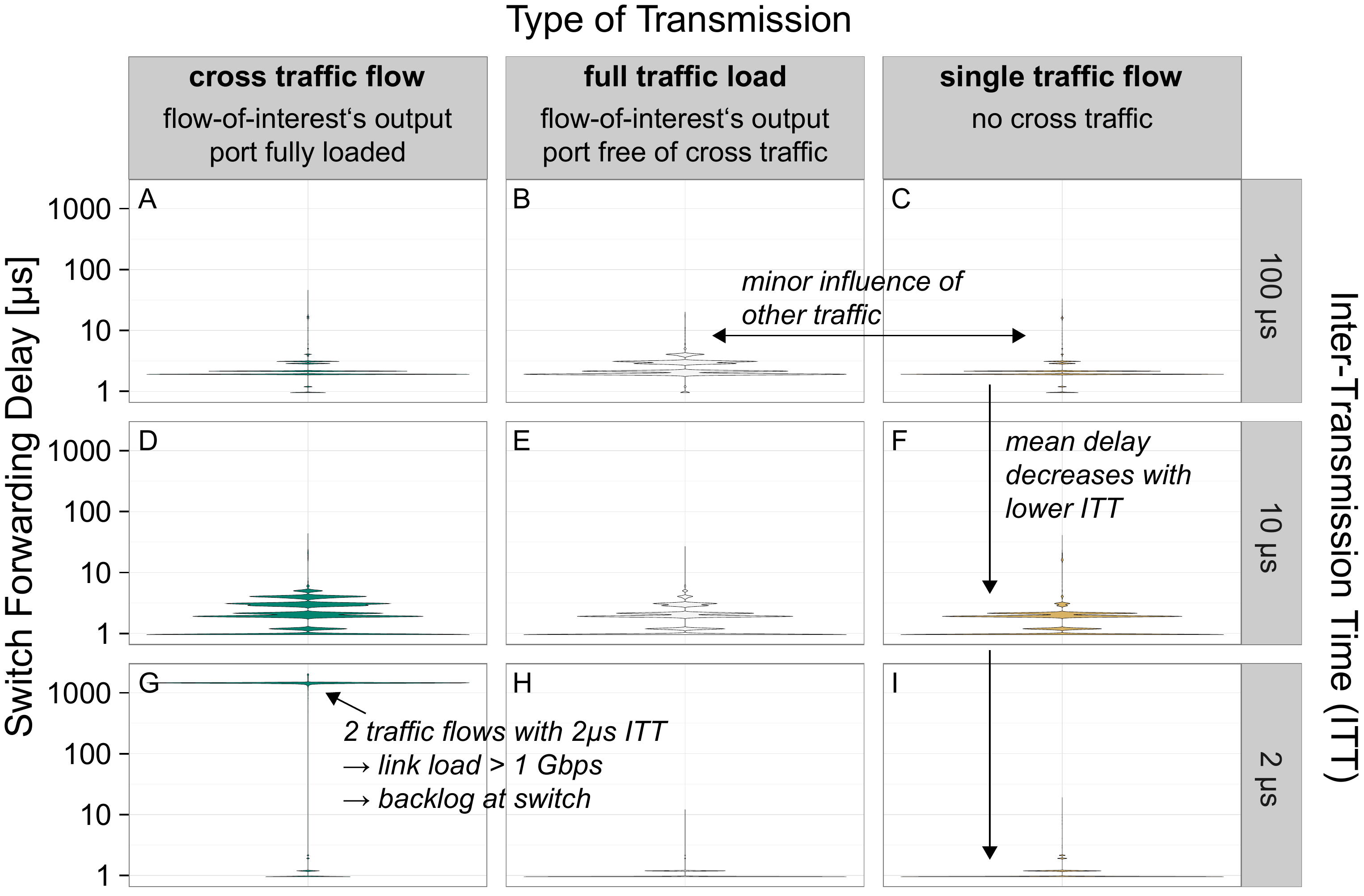}
	\caption{Traffic flow switching delay for different \acf{ITT} and (cross) traffic conditions on a \SI{1}{Gbps} network}
	\label{fig:nc_prep} 
\end{figure}

Preliminary studies for \ac{NC} application include the analysis of switching delays of a virtual switch for different \ac{ITT} and traffic conditions, as illustrated in Figure \ref{fig:nc_prep}.
It needs to be stressed that these evaluations only serve for confirming the assumptions on cross traffic behavior described in Section \ref{sec:approach:nc}.
They do not reflect actual traffic configurations considered in the remainder of this article.

To deduce latencies, traffic captures of one specific flow-of-interest at the ingress and egress port of the switch are considered.
The single traffic flow case constitutes a scenario, in which only the flow-of-interest is present, whereas in the cross traffic case a second flow uses the same egress port.
The full traffic load scenario involves additional communication streams, reaching the switch, however obviating the egress port used by the flow-of-interest.
It can be observed that the delay decreases with reduced \ac{ITT} (for a more detailed analysis of this phenomenon c.f. \cite{Kurtz2016empirical}).
Meanwhile, additional traffic at the switch shows minimal influence on the switching performance, if different egress ports are used.
In comparison, cross traffic being present on the same egress port, evokes rising delays of the flow-of-interest.
If the competing traffic flows exceed the maximum capacity of the connected egress link -- which is true for an \ac{ITT} of \SI{2}{\mu s} -- delay even increases by three orders of magnitude (c.f. Figure \ref{fig:nc_prep}.G). 
Accordingly, traffic using the same output port as the flow-of-interest needs to be considered for delay analysis due to its significant impact, whereas the influence of traffic flows on other output ports has been shown to be negligible.
Hence, \ac{NC} can be simplified in this regard, as described in Section \ref{sec:approach:nc}.
This obviates the issue of looped flow dependencies, which otherwise might cause deadlocks in computation \cite{yang2017analyzing}.
On the other hand, measurements reveal the need for considering the impact of varying \ac{ITT} on switching latencies.
Subsequently, these findings are integrated into \ac{NC}.

\subsubsection{Validation of Network Calculus Delay Bounds}
\label{sec:eval:nc:val}
\begin{figure}[b!]
	\centering
	\includegraphics[width=\columnwidth]{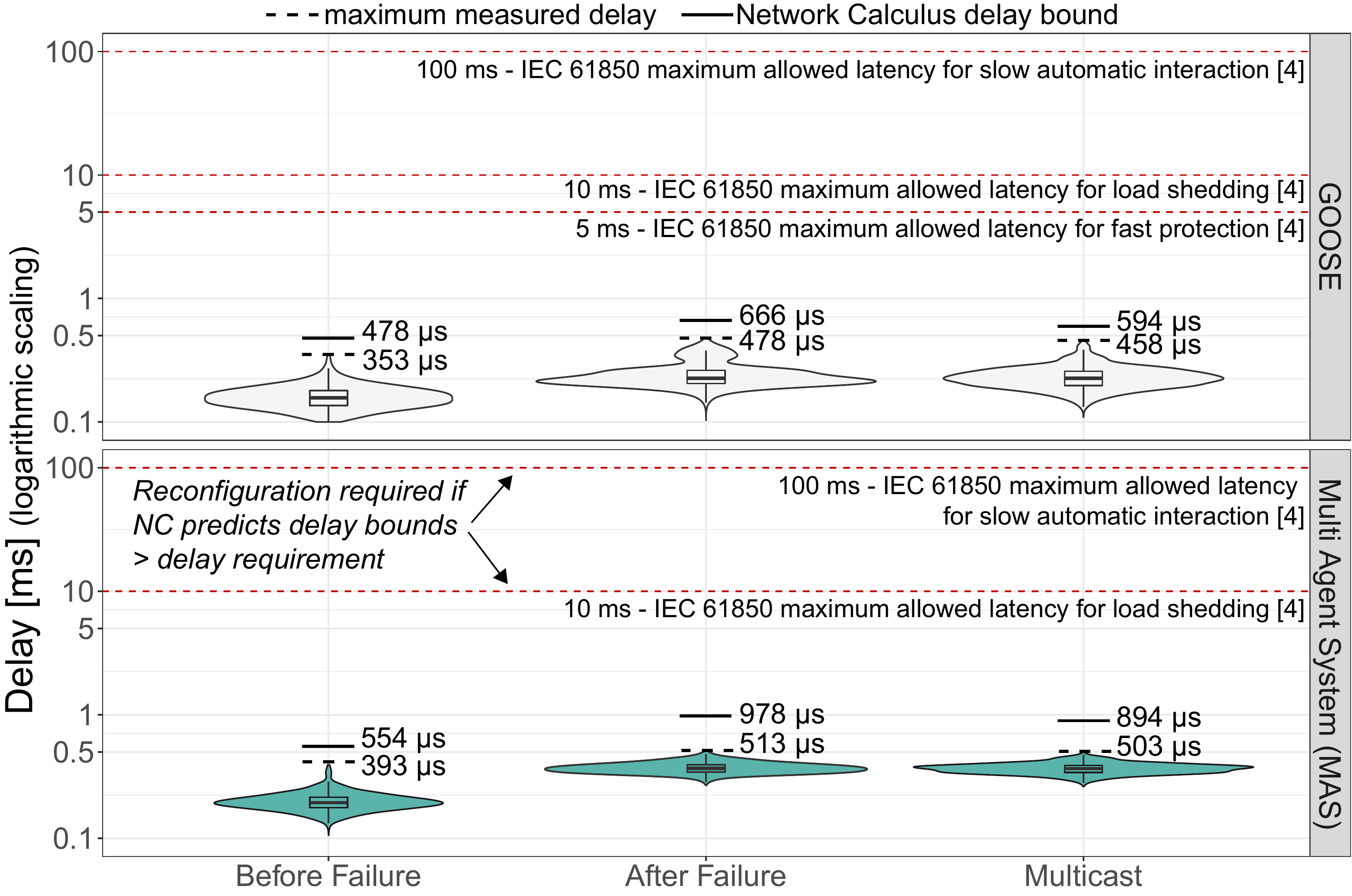}
	\caption{Measured delays (violin plots, box plots, dashed lines) and \acf{NC} bounds (solid lines) of \ac{GOOSE} and \acf{MAS} traffic from Substation 38 to 41 for different scenarios}
	\label{fig:nc_delays} 
\end{figure}

In the next step, we aim at comparing measured network delays to the results of \ac{NC}-based flow analysis in order to prove its applicability for network state monitoring and delay supervision.
Figure \ref{fig:nc_delays} comprises measured delays in terms of violin and box plots for \ac{GOOSE} and \ac{MAS} transmissions between the control center (Substation 38) and Substation 41, considering three different scenarios.
Above the violins, dotted lines indicate the maximum measured delay, whereas solid lines represent the corresponding results of in-controller \ac{NC} analysis.
In comparison to the previous evaluation, the traffic loads listed in Table \ref{tab:trafficpat} are restored.
Hence, the two flows-of-interest are interfered by multiple cross traffic flows.

The scenarios considered map to the use cases presented in the course of this paper: \textit{before failure} of the communication link between Substations 38 and 39, \textit{after failure recovery} to alternative paths and \textit{after applying multicast} transmission mode.
Dynamic prioritization is excluded here, since it would involve overloading communication links, resulting in infinite delay bounds in \ac{NC}.
      
In all three scenarios, \ac{NC} bounds are not exceeded, being $120$ to \SI{450}{\mu s} above the maximum values, measured in the testbed.
Deviations between \ac{NC} bounds and maximum measured values increase for the case of \ac{MAS} traffic after occurrence of the \ac{ICT} failure.
This effect can be attributed to \ac{NC}'s sensitivity to prioritization.
In this case, the behavior is sparked by relatively low priority of the \ac{MAS} service in combination with numerous -- higher priority -- cross traffic flows, being present on the back-up route.
Nevertheless, evaluation highlights that \ac{NC} provides valid means of network latency estimation within \acs{SUCCESS}.
Delay bounds are found to be well-above maximum measurement results, while not being overly loose.
Yet, it needs to be kept in mind that real-world systems might be extremely dynamic, experiencing sudden, unforeseen changes in delay or available bandwidth.
Unfortunately, \ac{NC} computation is not able to account for such situations directly.
However, there are two approaches to handle this challenge:
\begin{itemize}
	\item Periodic measurements can be used to ensure the validity of service and arrival curve models, as described in Section \ref{sec:approach:nc:model}.
	Yet, reasonable update intervals -- considering the induced additional network load -- might not be sufficient to handle sudden events.
	\item Due to its pessimistic nature (i.e. being based on worst case assumptions \cite{fidler_nc}), \ac{NC} includes a certain degree of tolerance against the impact of unforeseen events.
	\item In addition, a threshold (c.f. Algorithm \ref{alg:nccheck}) is introduced to ensure timely controller intervention.
	Thus, actions are taken before \ac{NC} delay bounds actually reach admissible delay requirements.
	In this way, the consequences of unforeseen factors can be compensated for.
	Here, we consider a threshold of \SI{10}{\%}.
	Measurements in real-world environments might be utilized to optimize this value.
\end{itemize}

In addition, the evaluations performed in this section provide an example of validating desired delay guarantees against the outcome of the established network configuration on basis of measurements. 
Additional comparisons were conducted for all flows in the scenario. 
However, this validation is performed offline. 
In an extension of our approach, such measures might be integrated in terms of a real-time feedback loop.

\subsubsection{Evaluation of Network Calculus-based Routing}
\label{sec:eval:nc:rout}

\begin{figure}[b!]
	\centering
	\includegraphics[width=\columnwidth]{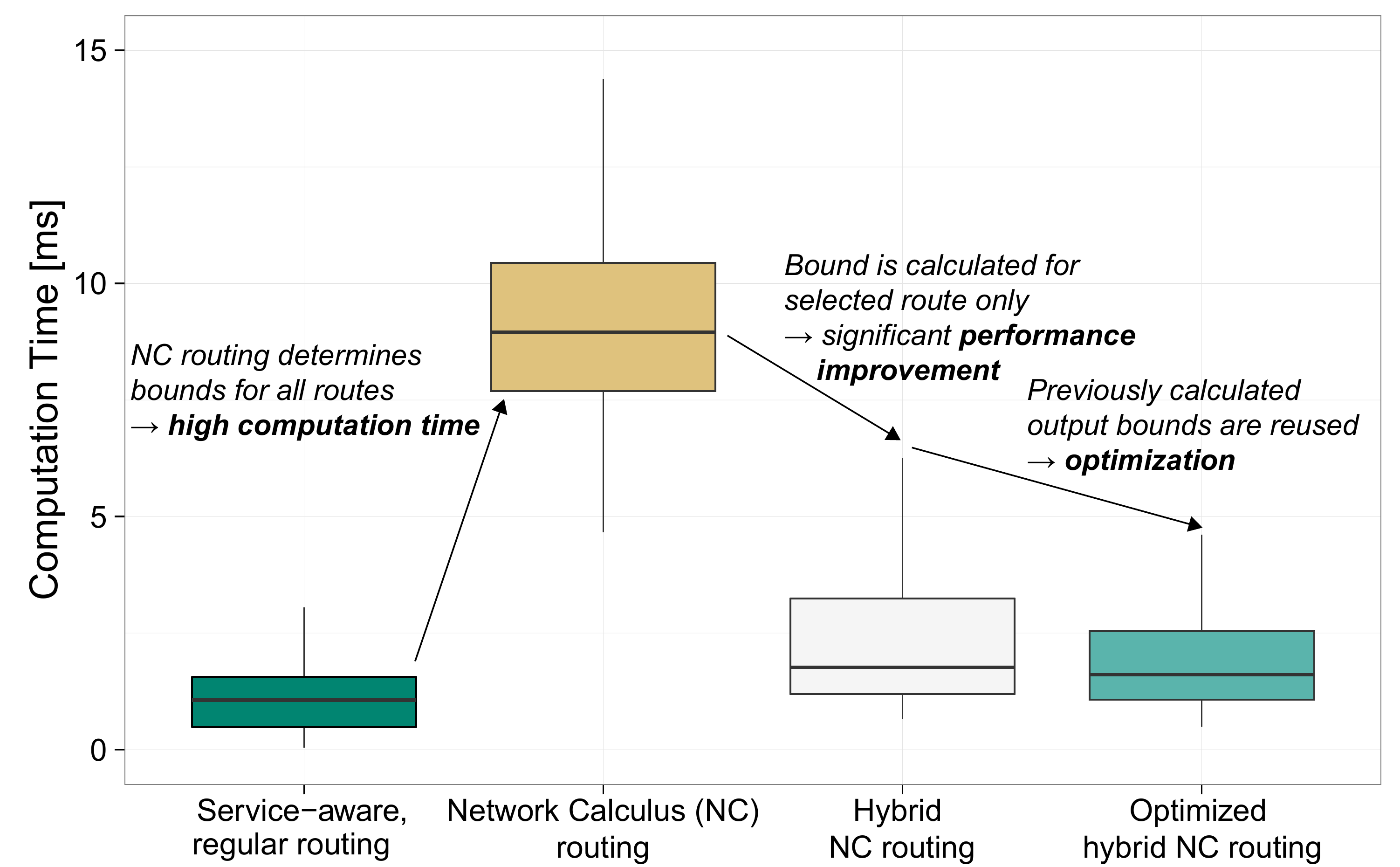}
	\caption{Comparison of computation times for regular, \acl{NC}-based and hybrid routing approaches, used in our \acf{SDN} Controller}
	\label{fig:nc_routing} 
\end{figure}

Figure \ref{fig:nc_routing} compares the performance of \ac{NC} based routing with the computation times of our regular, service-aware routing approach.  
While the regular routing completes within less than \SI{3}{ms} at maximum, full \ac{NC}-based routing incurs mean delays of \SI{14.44}{ms}.
Computation speed of this \ac{NC} routing approach is determined by the fact, that delay bounds are derived for all feasible routes within the full Nordic 32 communication network.
The performance of our algorithm might be improved by parallelizing calculations, e.g. assessing different routes simultaneously.

In contrast, the hybrid NC routing concept builds on the idea of coupling service-aware routing and \ac{NC} analysis.
Therefore, an optimal route is determined using regular routing, for which delay bound compliance is checked with the help of \ac{NC}.
Hence, performance is improved to mean computation times of \SI{2.66}{ms}.
To further optimize computation times of \ac{NC} routing, we re-use previously calculated output bounds during delay bound calculation for the new flow-of-interest as described in Algorithm \ref{alg:nccheck}.
This obviates efforts of recursively determining output bounds on-the-fly.
Subsequently, the mean calculation period is decreased to \SI{2.17}{ms} in case of optimized hybrid \ac{NC} routing, however at the cost of reduced precision of the delay bound.

\subsubsection{Optimization of Network Calculus Computation Times}
\label{sec:nc:eval:opt}

\begin{figure}[t]
	\centering
	\includegraphics[width=.95\columnwidth]{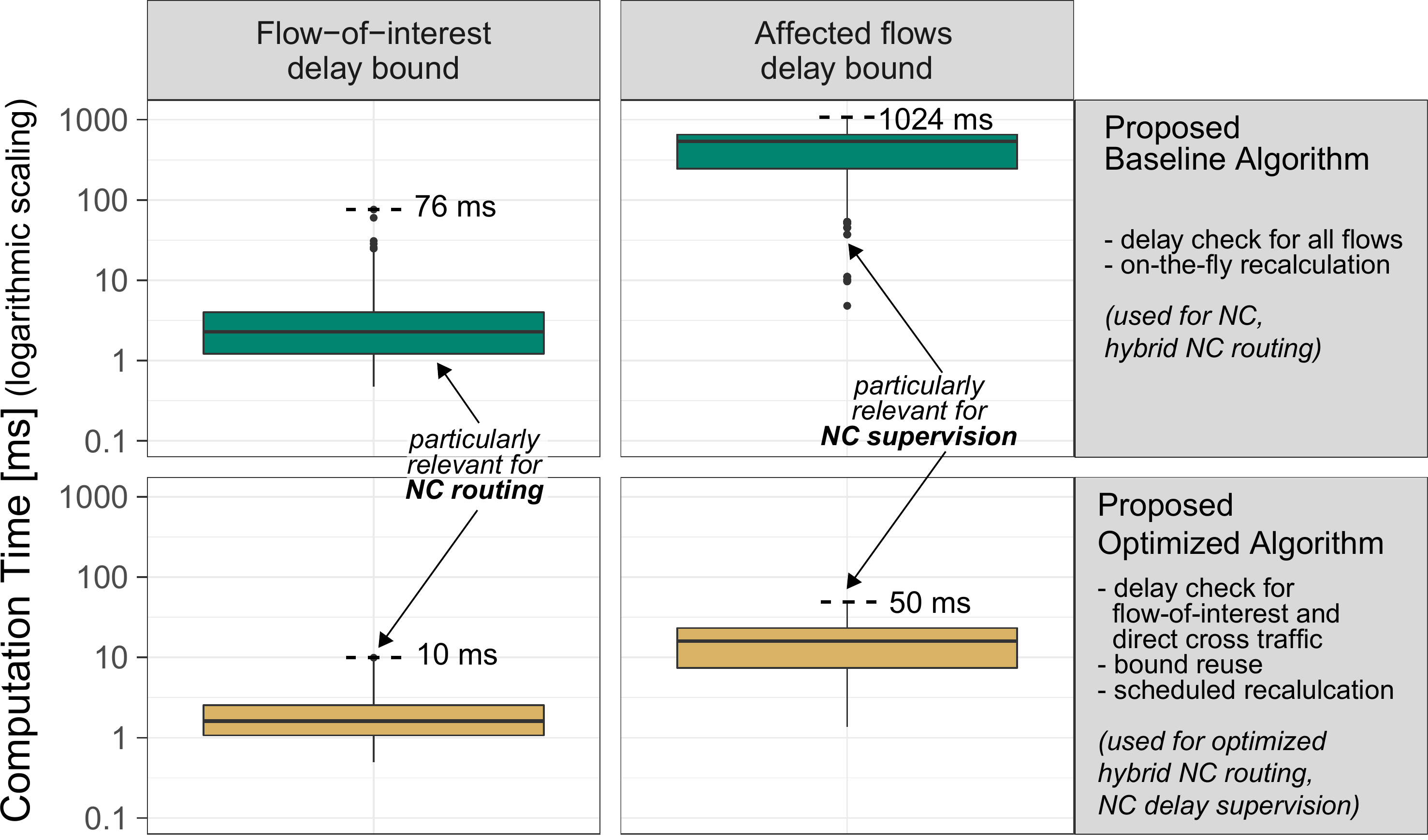}
	\caption{Comparison of computations times for different calculation objects and algorithms with relevant parameters for \ac{NC} routing respectively \ac{NC} delay supervision being highlighted}
	\label{fig:sdn_scen_nc_opt} 
\end{figure}

The following evaluation focuses on the optimization of \ac{NC} computation times for the application within the \ac{SDN} controller.
The performance of the baseline algorithm and the optimized approach are compared in Figure \ref{fig:sdn_scen_nc_opt}, displaying measured computation times for the complete Nordic 32 system.  
The baseline algorithm was utilized for \ac{NC} and hybrid \ac{NC} routing, whereas the enhanced version has been employed for optimized hybrid \ac{NC} routing as well as for \ac{NC} delay supervision.
Following the baseline approach, output bounds of all cross traffic flows are computed on-the-fly during delay calculation of the flow-of-interest (first column).
This leads to maximum computation times of \SI{76}{ms}.
Afterwards, the delay of all previously installed traffic flows is recalculated, considering the impact of the new flow (second column).
This step may take up to approximately \SI{1}{s}.

Initial delay analysis of the flow-of-interest can be sped up by making use of previously calculated output bounds.
Thus, calculation times can be reduced to maxima of \SI{10}{ms} for the flow-of-interest and \SI{50}{ms} for affected cross traffic flows.
The latter provides a worst-case estimation as delay bounds for all cross traffic flows are recomputed.
In real-world scenarios it would be sufficient to recalculate the delay bounds of those flows close to their respective latency requirements.
Due to the concept of reusing existing output bounds, it becomes necessary to perform a third calculation step, recalculating the output bounds.
Nevertheless, this final step does not need to be executed immediately, but may be scheduled.

\begin{figure}[t]
	\centering
	\includegraphics[width=.95\columnwidth]{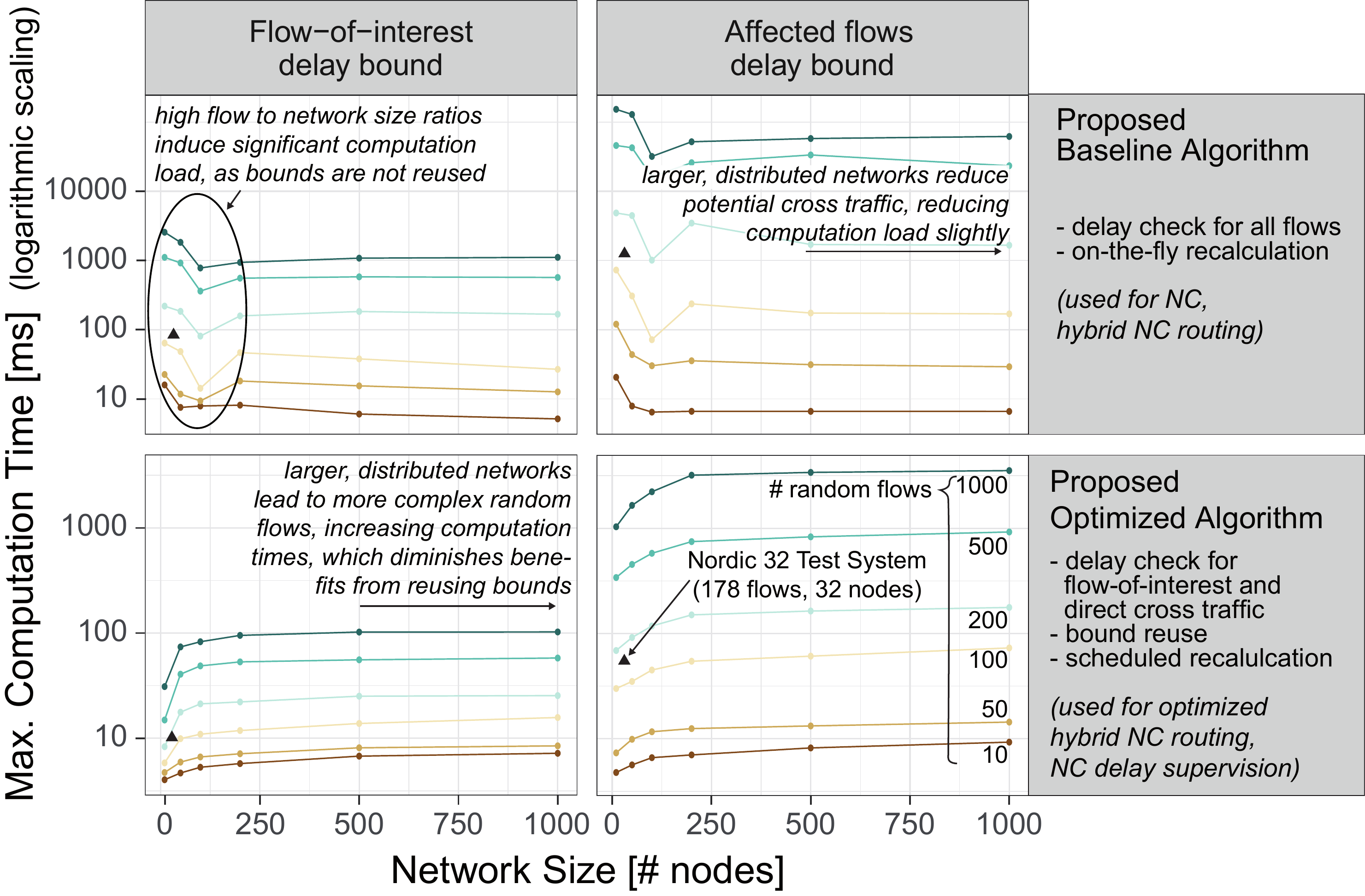}
	\caption{Scalability of \ac{NC} algorithms, integrated into the \ac{SDN} controller, with regard to computation times, when varying network sizes and numbers of flows}
	\label{fig:sdn_scen_nc_opt_scale} 
\end{figure}

This evaluation is complemented by the scalability analyses, provided in Figure \ref{fig:sdn_scen_nc_opt_scale}.
For this purpose, maximum computation times of the two proposed algorithms are displayed for both applications, i.e. routing and delay supervision.
On the x-axis network size is varied in terms of increasing numbers of interconnected nodes.
In the previous scenarios, we applied a realistic communication network topology based on the Nordic 32 reference power system.
However, for investigating scalability, we utilize the Barab\'{a}si-Albert model \cite{barabasi_albert} to generate random graph topologies. 
Based on these network scenarios, rising numbers of random traffic flows are created, illustrated by the sets of curves in Figure \ref{fig:sdn_scen_nc_opt_scale}.
To obtain adequate results, the evaluations are performed for $100$ different seeds of the random number generator, providing different topologies and flow configurations.
Each of the four fields in Figure \ref{fig:sdn_scen_nc_opt_scale} contains a triangle symbol, which represents the corresponding results of the Nordic 32 system.

Similar to the evaluations in Figure \ref{fig:sdn_scen_nc_opt}, it is apparent that the proposed optimized algorithm outperforms the respective baseline approach.
For example, delay bound calculations for the flow-of-interest in \ac{NC} routing may require up to approximately \SI{1}{s} (1000 flows, 1000 network nodes), when applying the proposed baseline algorithm.
Using the optimized approach, computation times can be reduced to about \SI{100}{ms} for the same configuration.
Overall, for all approaches and applications, computation times increase with rising numbers of considered traffic flows.

However, with regard to network size, the curves of the two algorithms indicate different scaling properties.
In case of the optimized algorithm, computation times experience logarithmic growth with increasing network size. 
The approach profits from very small networks with several flows sharing the same paths.
Thus, the gain from reusing previously calculated bounds is maximized.
By extending the topology, the advantage declines as the random flows become ever more complex, leading to significantly higher computation times.
Nevertheless, when the network size is further increased this effect is balanced, as flows are less likely to interfere.
Hence, the rise of computation times is weakened.

In contrast, small network topologies can be seen as a worst case scenario for the proposed standardized algorithm.
In such systems, especially under high loads, interference between traffic flows is maximized.
Similar delay bounds have to be computed repeatedly, as there is no re-use of existing bounds.
Subsequently, computation times drop with increasing network sizes due to reduced interference.
Though, when the topology is further extended, similar effects as for the optimized approach apply.
Thus, computation times experience another rise.
However, for very large systems, the balance between the different effects shifts.
Enhanced distribution of traffic flows among the network leads to slight reductions of computational loads.

Besides comparing \ac{NC} algorithms, Figure \ref{fig:sdn_scen_nc_opt_scale} points out limitations of our proposed routing and delay supervision concepts.
To comply with \ac{IEC}~61850 service requirements, the area supervised by a single controller needs to be confined to a certain combination of network nodes and flows.
For example, up to about 100 flows may be managed on topologies of up to 1000 nodes.
In contrast, orchestrating 200 transmissions requires restricting the network to about 50 nodes.
This investigation is continued in the following section.

\subsubsection{Assessment of Delay Supervision for Dynamic Reconfiguration}

\begin{figure}[t]
	\centering
	\subfloat[Nordic 32 system measurements]{
		\includegraphics[width=0.38\columnwidth]{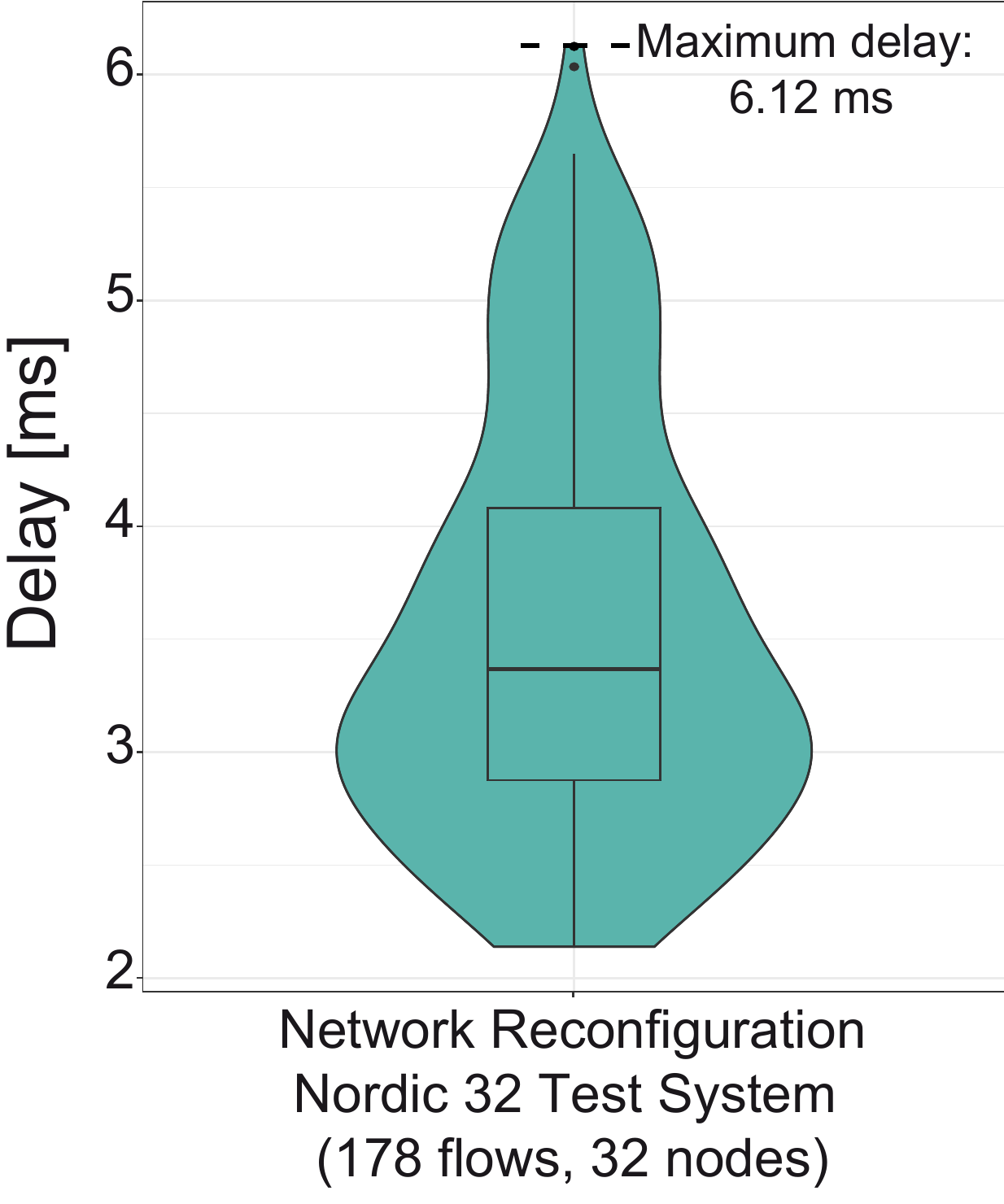}
		\label{fig:reconfig_delay} 
	}
	\subfloat[Scalability analysis using random topologies, flows]{
		\includegraphics[width=0.57\columnwidth]{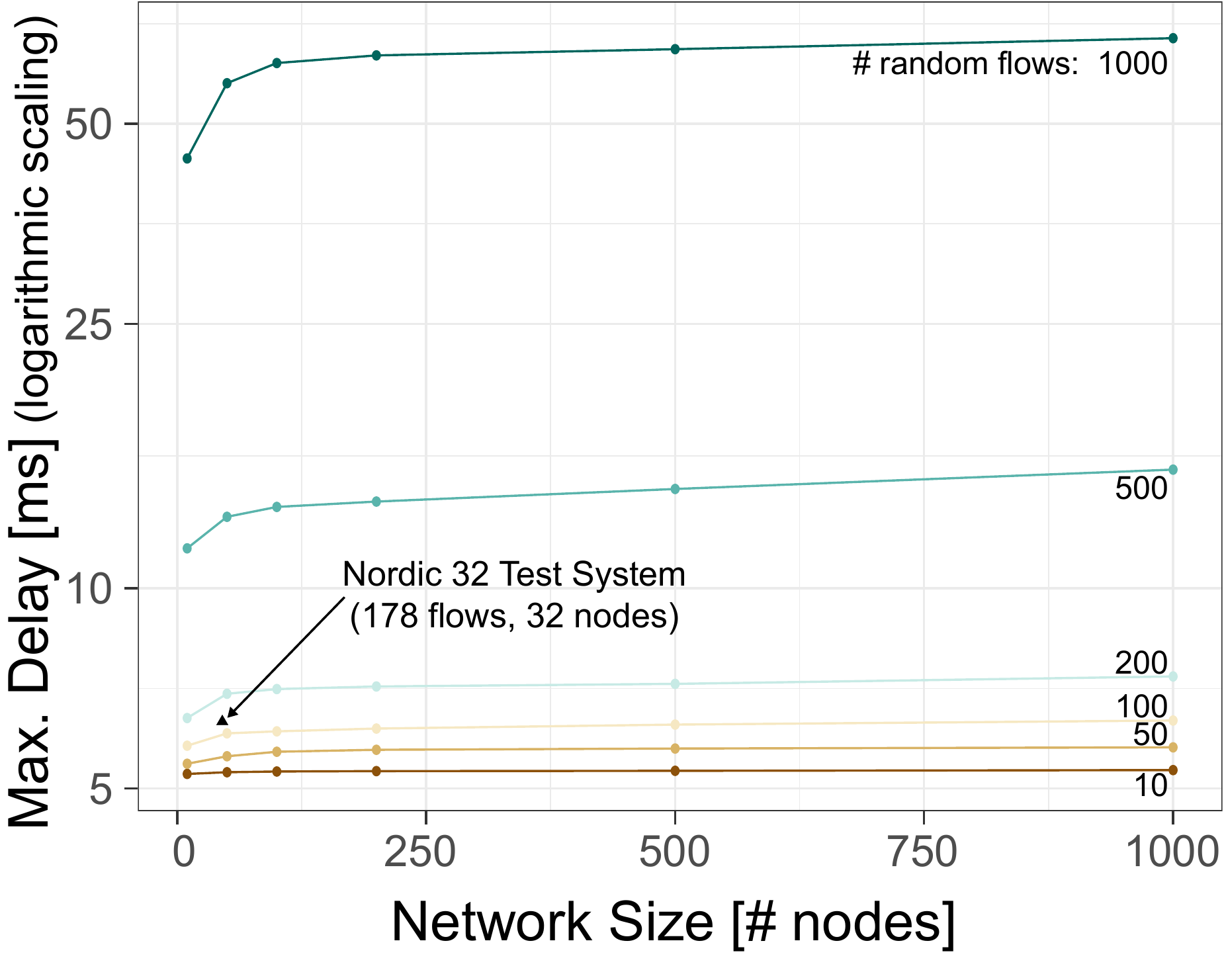}
		\label{fig:reconfig_delay_scale} 
	}
	\caption{Delay incurred by network reconfiguration}
	\label{fig:reconfig_delay_all}
\end{figure}

\definecolor{LightGray}{cmyk}{0,0,0,0.17}
\newcolumntype{g}{>{\columncolor{LightGray}}c}
\begin{table}[t]
	\footnotesize
	\centering
	\caption{Delay impact of computation times  derived from the results presented in Figures \ref{fig:sdn_scen_nc_opt}, \ref{fig:sdn_scen_nc_opt_scale} and \ref{fig:reconfig_delay_all}}
	\begin{tabular}{l c c c c c c c }
		\toprule
		\multirow{3}{2.2cm}{\textbf{Options}} & \multirow{3}{2.2cm}{\textbf{Chain of Events}} & \multicolumn{6}{c}{\textbf{Max. Delay impact [ms]}}\\
		\addlinespace[1mm]
		\cline{3-8}
		\addlinespace[1mm]
		& & \multicolumn{3}{c}{Request. flow} & \multicolumn{3}{c}{Affected flows} \\
		\addlinespace[1mm]
		\cline{3-8}
		\addlinespace[1mm]
		& \multicolumn{1}{r}{\parbox{7mm}{Flows\\Nodes}}& \parbox{5mm}{178$^*$\\32} & \parbox{5mm}{100\\1000} & \parbox{5mm}{200\\100} & \parbox{5mm}{178$^*$\\32} & \parbox{5mm}{100\\1000} & \parbox{5mm}{200\\100} \\
		\midrule
		\multirow{6}{0.7cm}{Post-reconfiguration check} & \parbox{2.8cm}{1. Request} & 6 & 6 & 6 & - & - & - \\
		\addlinespace[1mm]
		\cline{2-8}
		\addlinespace[1mm]
		& \parbox{2.8cm}{2. Reconfiguration of \\    requesting flow} & 6 & 6 & 6 & - & - & - \\
		\addlinespace[1mm]
		\cline{2-8}
		\addlinespace[1mm]
		& \parbox{2.8cm}{3. NC recalculation} & - & - & - & 49 & 72 & 92 \\
		\addlinespace[1mm]
		\cline{2-8}
		\addlinespace[1mm]
		& \parbox{2.8cm}{4. Reconfiguration of\\ affected flows} & - & - & - & 6 & 6 & 9\\
		\addlinespace[1mm]
		\cline{2-8}
		\addlinespace[1mm]
		& \parbox{2.8cm}{\textbf{Total}} & \cellcolor{Gray}\textbf{12} & \cellcolor{Gray}\textbf{12} & \cellcolor{Gray}\textbf{12} & \cellcolor{Gray}\textbf{56} & \cellcolor{Gray}\textbf{78} & \cellcolor{Gray}\textbf{101}\\
		\midrule
		\multicolumn{8}{l}{\parbox{11cm}{$\rightarrow$ in the worst case, \textbf{affected flows} impacted considerably}}\\
		\midrule
		\multirow{6}{2.2cm}{Pre-reconfiguration check} & \parbox{2.8cm}{1. Request} & 6 & 6 & 6 & - & - & - \\
		\addlinespace[1mm]
		\cline{2-8}
		\addlinespace[1mm]
		& \parbox{2.8cm}{2. NC recalculation} & 49 & 72 & 92 & - & - & - \\
		\addlinespace[1mm]
		\cline{2-8}
		\addlinespace[1mm]
		& \parbox{2.8cm}{3. Reconfiguration of\\ affected flows} & 6 & 6 & 9 & - & - & - \\
		\addlinespace[1mm]
		\cline{2-8}
		\addlinespace[1mm]
		& \parbox{2.8cm}{4. Reconfiguration of\\requesting flow} & 6 & 6 & 6 & - & - & - \\
		\addlinespace[1mm]
		\cline{2-8}
		\addlinespace[1mm]
		& \parbox{2.8cm}{\textbf{Total}} & \cellcolor{Gray}\textbf{68} & \cellcolor{Gray}\textbf{90} & \cellcolor{Gray}\textbf{113} & \cellcolor{Gray}\textbf{0} & \cellcolor{Gray}\textbf{0} & \cellcolor{Gray}\textbf{0} \\
		\midrule
		\multicolumn{8}{l}{\parbox{11cm}{$\rightarrow$ in the worst case \textbf{requesting flow} impacted considerably}}\\
		\midrule
		\midrule
		\multicolumn{8}{l}{\parbox{11cm}{$\rightarrow$ applicable for \textbf{Smart Grid} services with \textbf{latency requirements} $\geq$\SI{100}{ms},\\assuming limited controller partitions}}\\
		\bottomrule
		\bottomrule
		\multicolumn{8}{l}{\parbox{11cm}{$^*$\textit{Nordic 32 reference system}}}\\
	\end{tabular}
	\label{tab:sdn_nc_reconfig_opt2}
\end{table}

Finally, the application of \ac{NC} delay supervision in the context of dynamic network reconfiguration is evaluated.
As shown in Figure \ref{fig:sdn_nc_int}, reconfiguration may be caused by the insertion of new traffic flows, as direct and indirect result of \ac{NBI} requests or evoked by failure recovery.
In this context, Figure \ref{fig:reconfig_delay} comprises measurement results for the delay of network reconfiguration in terms of a violin and overlaid box plot.
The median reconfiguration time amounts to \SI{3.37}{ms}, whereas at maximum delays of \SI{6.12}{ms} are reached.

Analogous to Figure \ref{fig:sdn_scen_nc_opt_scale}, Figure \ref{fig:reconfig_delay_scale} assesses scalability in terms of maximum reconfiguration times, depending on network size and number of flows.
Supporting the results of the previous evaluations (c.f. Section \ref{sec:nc:eval:opt}), it is shown that the number of flows is a particularly limiting factor for dynamic network reconfiguration.
In comparison, the impact of network size is minor.
Considering \ac{IEC}~61850 latency requirements, the reconfiguration of up to 200 flows is regarded as manageable.
The obtained reconfiguration times are taken into account for subsequent analyses.

Table \ref{tab:sdn_nc_reconfig_opt2} focuses on the case of \ac{NBI} request-induced network reconfiguration, comparing delay impact of different implementation options.
These alternatives deviate with regard to the order, in which processes are executed.
In case of \textit{post-reconfiguration check} the network configuration (queue rate, priority) is altered immediately, resulting in maximum adjustment latencies of about \SI{12}{ms} for the requesting flow in the Nordic 32 reference system.
Only afterwards, \ac{NC} is employed to recalculate the delay bounds of affected flows and check for potential violations of given latency requirements.
If so, subsequent reconfiguration of the affected traffic flows has to be performed.
Accumulating \ac{NC} computation and corresponding reconfiguration times, a worst case delay of \SI{56}{ms} is constituted.

In contrast, using the \textit{pre-reconfiguration check} other flows are not influenced by the \ac{NBI} request as potential effects on their delay bounds are assessed beforehand.
However, in this way the reconfiguration of the requesting flow is delayed by up to \SI{68}{ms} in the Nordic 32 system.
Hence, both approaches exhibit advantages and disadvantages, either for the requesting flow or for affected transmissions.
Further, Table \ref{tab:sdn_nc_reconfig_opt2} comprises two additional network and flow configurations taken from the evaluations in Figures \ref{fig:sdn_scen_nc_opt_scale} and \ref{fig:reconfig_delay_scale}.
The second parameter set (100 flows on 1000 nodes) allows reconfiguration times just below \SI{100}{ms}, whereas the third (200 flows on 50 nodes) yields latencies slightly above this value.
Taking into account Smart Grid latency requirements defined in Table \ref{tab:timereq} as well as the different network configurations investigated (c.f. Figures \ref{fig:sdn_scen_nc_opt_scale} and \ref{fig:reconfig_delay_scale}), the following conclusions can be drawn:
\begin{itemize}
	\item Combining \ac{NC} delay supervision with dynamic network reconfiguration allows 
	for flexibly reallocating resources for Smart Grid traffic flows with latency requirements $\geq$\SI{100}{ms} as delay compliance is ensured at all times.
	However, the network partition supervised by a single controller needs to be limited in size and number of flows. 
	Feasible extrema of configuration are the following: up to 100 flows and 1000 nodes or up to 200 flows and 10 nodes.
	Besides, there are further possible combinations in between.
	\item In contrast, extremely time critical services with latency requirements $<$\SI{10}{ms} may not be subjected to reconfiguration at any time.
	\item Vice versa, minimum and maximum queue concepts have to be employed for assuring dedicated resources for these services.
	Respective configurations must not be altered during failover or reconfiguration.
	\item Further optimization of algorithms and hardware set-up may enable extending dynamic, \ac{NC} monitored network reconfiguration to Smart Grid services with latency requirements of $10$-\SI{100}{ms}.
	Currently, feasible network configurations range from 10 flows and 200 nodes to 50 flows and 10 nodes.
\end{itemize}
Overall, the evaluation results highlight that applicability and performance of \ac{NC} routing and delay supervision are tightly coupled to the dimensioning of network partitions, i.e. the areas orchestrated by one controller. 
At the same time, these interdependencies raise the issue of coordinating \ac{NC} operations between multiple controllers.

\section{Conclusion and Future Work}
\label{sec:conclusion}
To cope with the complex challenges of mission critical communications in cyber-physical systems, we proposed the use of \acf{SDN} on basis of our \acf{SUCCESS} framework.
In this article we focused on the case of emerging Smart Grid infrastructures, evaluating the suitability of our approach with the help of experiments and emulations.
Therefore we modeled an \ac{ICT} infrastructure on top of the well-established Nordic 32 test system and derived specific scenarios for each aspect of hard service guarantees.

Reliability of communication networks was studied with regard to handling critical link failures.
Applying a hybrid concept, combining distributed and centralized failure detection and recovery, maximum delays of \SI{5}{ms} are achieved, while maintaining optimal paths almost continuously.

Dynamic adaptation of priorities (queues) is utilized for minimizing communication delays of a \acf{MAS}, even in the presence of high traffic load.
Alternating requirements are conveyed via the controller's \acf{NBI}, relying on the \ac{REST} \ac{API}.
In addition, the \ac{NBI} is used for creating multicast groups, as commonly used in \ac{IEC}~61850 communications, significantly reducing average and maximum link load.

Finally, the analytical modeling approach of \acf{NC} was integrated into \acs{SUCCESS} and tailored to the specifics of min/max rate queuing as implemented at the switches within our testing environment.
Hence, real-time capability of critical communications can be monitored online on basis of hard worst case delay bounds.
In case of violations, remedial actions, such as fast re-routing or dynamic priority adaptation, are applied.
In contrast to measurement-based latency supervision, \ac{NC} integration enables a comprehensive view on delays, their triggers and even predictions of future endangerments.
Yet, we also indicated limits of \ac{NC}-monitored dynamic network reconfiguration as -- for numerous traffic flows -- computation times may jeopardize latency requirements of extremely time critical Smart Grid protection functions ($<$\SI{10}{ms}).
Further, \ac{NC} was utilized for improved, delay-bounded routing.

Further enhancing our reliability concept, subsequent work will deal with fast failure recovery for multicast traffic flows.
Moreover, we aim at establishing communication between distributed, inter-connected controllers in order to achieve a) controller resilience and b) improve the scalability.
With respect to the latter, the realization of \ac{NC}-enabled routing and delay supervision in infrastructures, with individual controllers for different network partitions, presents an interesting field of further research.
Major challenges include the handling of traffic flows, traversing multiple controller domains.
Additionally, assignment of transmission capacities in wireless networks can be added to the controller's capabilities.

\section*{Acknowledgement}
\small
This work has been carried out in the course of research unit 1511 \textit{'Protection and control systems for reliable and secure operations of electrical transmission systems'}, funded by the German Research Foundation (DFG) and the Franco-German Project \textit{BERCOM} (FKZ: 13N13741) co-funded by the German Federal Ministry of Education and Research (BMBF).

\bibliography{sdnbib}

\end{document}